\documentclass[a4paper]{article}
\usepackage[english]{babel}
\usepackage[T2A]{fontenc}
\usepackage[utf8]{inputenc}
\usepackage{amsmath}
\usepackage{amsthm}
\usepackage{amssymb}
\usepackage{textcomp}
\usepackage{graphicx}
\usepackage[unicode,pdfborder={0 0 1}]{hyperref}

\usepackage{todonotes}

\newcommand{\eqlb}[2]{\begin{equation} \label{#1} #2 \end{equation}}
\newcommand{\eq}[1]{\begin{equation} #1 \end{equation}}
\newcommand{\eqn}[1]{\begin{equation*} #1 \end{equation*}}

\newcommand{\eqs}[1]{$#1$}
\newcommand{\brc}[1]{\left(#1\right)}
\newcommand{\bsq}[1]{\left[#1\right]}
\newcommand{\bfi}[1]{\left\{ #1\right\}}

\newcommand{\abs}[1]{\left|#1\right|}

\newcommand{\qq}{\qquad}
\newcommand{\ds}{\displaystyle}

\newcommand{\matr}[2]{\begin{array}{#1}#2\end{array}}

\newcommand{\at}[2]{\genfrac{}{}{0pt}{}{#1}{#2}}

\newcommand{\rme}{\textrm{e}}
\newcommand{\rmd}{\textrm{d}}

\newcommand{\sn}{\textrm{sn}}

\newcommand{\tc}{\tilde{c}}
\newcommand{\wc}{\widehat{C}}

\numberwithin{equation}{section}

\hoffset= - 1.0in
\begin{document}
\title{{\bf {New non-linear equations and modular form expansion for double-elliptic Seiberg-Witten prepotential} \vspace{.2cm}}
\author{{\bf G. Aminov$^{a,b,}$}\footnote{aminov@itep.ru}, {\bf A. Mironov$^{c,a,d,e,}$}\footnote{mironov@itep.ru; mironov@lpi.ru}  \ and  {\bf A. Morozov$^{a,d,e,}$}\thanks{morozov@itep.ru}}
\date{ }
}

\maketitle

\vspace{-6.0cm}

\begin{center}
\hfill FIAN/TD-14/16\\
\hfill IITP/TH-11/16\\
\hfill ITEP/TH-14/16
\end{center}

\vspace{4.2cm}

\begin{center}
$^a$ {\small {\it ITEP, Moscow 117218, Russia}}\\
$^b$ {\small {\it Moscow Institute of Physics and Technology, Dolgoprudny 141700, Russia }}\\
$^c$ {\small {\it Lebedev Physics Institute, Moscow 119991, Russia}}\\
$^d$ {\small {\it National Research Nuclear University MEPhI, Moscow 115409, Russia }}\\
$^e$ {\small {\it Institute for Information Transmission Problems, Moscow 127994, Russia}}
\end{center}

\vspace{1cm}

\begin{abstract}
Integrable $N$-particle systems have an important property that the
associated Seiberg-Witten prepotentials satisfy the WDVV equations.
However, this does not apply to the most interesting class of
elliptic and double-elliptic systems.
Studying the commutativity conjecture for theta-functions on the families of
associated spectral curves, we derive some other non-linear
equations for the perturbative Seiberg-Witten prepotential,
which turn out to have exactly the double-elliptic system as their generic solution.
In contrast with the WDVV equations, the new equations acquire non-perturbative corrections
which are straightforwardly deducible from the commutativity conditions.
We obtain such corrections in the first non-trivial case of $N=3$ and describe
the structure of non-perturbative solutions as expansions in powers of the flat moduli
with coefficients that are (quasi)modular forms of the elliptic parameter.
\end{abstract}

\section{Introduction}

Seiberg-Witten (SW) theory \cite{SW1,SW2} is a foundation of many branches of modern theory.
It is a quasiclassical limit with respect to the peculiar $\epsilon$-variables of Nekrasov theory,
which, on one side, is AGT related to the two-dimensional conformal theory, Chern-Simons and knot theories,
and, on another side, is linked to a combinatorics of 3d partitions, tropical
geometry of Calabi-Yau spaces, refined topological vertices, all these being described in terms
of various matrix models and $\beta$-ensemble.
Remarkably, Seiberg-Witten theory {\it per se} is just equivalent to the theory
of integrable $N$-particle systems \cite{GKMMM},
thus all the above subjects should and do possess an interpretation in basic
terms of group theory.
This equivalence, however, implies and requires an extension of the well-known
set of integrable models of the Calogero-Ruijsenaars type to include their duals
\cite{Ruijs,Ruj'95,FGNR,MM,GM} and, most important, the self-dual  {\it double-elliptic}
integrable system, of which just the very initial facts are already  known  \cite{BMMM'2000,MM,MM2,AMMZ,ABMMZ}.
(There is another approach based on spectral dualities \cite{MTV,MTV2,BPTY,MMZZ13,MMRZZ13,MMRZZ34}, for the last important development in {\it this}
direction, also relevant for the double-elliptic case, see \cite{MMZ'16}.)

Seiberg-Witten theory interprets the eigenvalues of Lax operator as a 1-form
on the spectral curve and treats integrals along the $A$-cycles as flat moduli $a_I$,
while those along the $B$-cycles as the gradient $\partial\mathcal{F}/\partial a_I$ of a function
$\mathcal{F}(a)$ known as Seiberg-Witten prepotential.
Such a description is possible due to symmetricity of the period matrix
$T_{IJ}=\partial^2\mathcal{F}/\partial a_I\partial a_J$.
From the point of view of Riemann surfaces,
the procedure works only for some peculiar Seiberg-Witten families,
namely, for those that are families of the spectral curves of integrable systems.
It is natural to ask for a more straightforward
definition in terms of the period matrix, not referring to the subtle question
of enumerating integrable systems.
At least, one could ask for some equation distinguishing the relevant functions
$\mathcal{F}\brc{a}$.
An attempt of this kind was made in \cite{MMM'96,MMM'97,MMM'00,MM'98,BMMM1'99}, where it was shown that
many SW prepotentials satisfy the "generalized" WDVV equations \cite{WitWDVV,DVVWDVV}
(these should not be mixed with the ones studied in \cite{Dub}: essential
for SW theory is absence of a distinguished modulus providing a constant metric).
The problem, however, was that "many" did not mean "all":
the most interesting elliptic systems (associated with the UV-finite
SUSY theories) did not fit into this class (see also \cite{BMMM'2007}).
Since then, the question of "an exhaustive equation" for the prepotentials
remains open.

In this paper, we make a considerable step towards resolving this
longstanding problem: we suggest an equation for the {\it perturbative}
prepotential, which is also non-linear, involve the third derivatives of the prepotential, but different from the generalized WDVV equation,
and which turns out to have the most general double-elliptic system
as its generic solution.
We also demonstrate how the {\it non-perturbative} corrections
are systematically built from this solution, but do not provide
a complete description of the equation for the full non-perturbative prepotential.
Also our very concrete formulas below are limited to the first essentially non-trivial example
of $N=3$ (three particles).

We obtain the equation by studying the conjecture of \cite{MM}
that the Poisson-commuting Hamiltonians can be made from theta-functions
on the SW families of Riemann surfaces,
with the Jacobian points $z_I$ and flat moduli $a_I$ playing the role of conjugate variables.
Since moduli appear in the theta-functions only through the period matrix $T_{IJ}$, the involutivity conditions for the Hamiltonians are
\eq{\bfi{H_a,H_b}=\sum_{i=1}^{N}\sum_{j\leqslant k}
\frac{\partial^3 \mathcal{F}}{\partial a_i\partial a_j\partial a_k}
\brc{\dfrac{\partial H_a}{\partial z_{i}}\dfrac{\partial H_b}{\partial T_{jk}}-
\dfrac{\partial H_b}{\partial z_{i}}\dfrac{\partial H_a}{\partial T_{jk}}}=0.}
These conditions can be simplified by rewriting them as $z$-independent relations, which was done in \cite{ABMMZ}, and the result is the following set of equations:
\eqlb{eq:eqPrep}{\sum_{i,j,k=1}^{N}\frac{\partial^3 \mathcal{F}}{\partial a_i\partial a_j\partial a_k} C^{\vec{\alpha}}_{ijk}=0, \quad \vec{\alpha}\in\mathbb{Z}^g/3\mathbb{Z}^g,}
where \eqs{C^{\vec{\alpha}}_{ijk}} are the theta constants of genus $g$ defined in (\ref{eq:thetacoord}) and $g=N$. We already discussed some simple properties of (\ref{eq:eqPrep}) in \cite{ABMMZ}, but now we show that these equations can be used to calculate the Seiberg-Witten prepotentials including their instanton corrections. In this sense, we propose the equations that follow from the integrability of SW theory.
We present the calculations for the three-particle ($N=3$) elliptic integrable systems associated with the low-energy limit of ${\cal N}=2$ SUSY gauge theories with adjoint matter hypermultiplets, the presentation being performed in the form allowing an immediate extension to an arbitrary number of particles $N$. These systems are: the elliptic Calogero-Moser system \cite{Calogero'75,Calogero'76,Moser'75,OP'81}, which is related to the $4d$ theory \cite{DW,IM'1996,IM'1997}, the elliptic Ruijsenaars system \cite{Ruj'87,Ruijs}, which is related to the $5d$ theory with one compactified Kaluza-Klein dimension \cite{BMMM'1999}, and the double-elliptic integrable system related to the $6d$ theory with two compactified Kaluza-Klein dimensions \cite{BMMM'2000,MM,MM2,Braden2003}.

We demonstrate that the non-perturbative prepotential is a series in flat moduli with the coefficients being modular forms. This fact is in a complete agreement with modular properties of the spectral curves of the corresponding integrable systems, and part of the behaviour (the dependence on the quasimodular form $E_2$) is described by the modular anomaly equation \cite{MNW} in the Calogero and the Ruijsenaars cases. In the Calogero case, it was revisited in the case of Nekrasov functions in \cite{BFL1,BFL2,BFL3,BFL4} and is associated with the modular transformations in $2d$ conformal field theories \cite{GMM,Nem}. In the double-elliptic case, it is more involved and will be discussed elsewhere \cite{AMM}.

\section{Involutivity conditions}
In this section, we recall the involutivity conditions obtained previously in \cite{AMMZ,ABMMZ}.
The $N$-particle Hamiltonians for the systems under consideration were constructed in \cite{BMMM'2000,MM} and can be represented as follows:
\eqlb{eq:dellH}{H_a=\dfrac{\theta\left[\begin{matrix}0\,\dots\,0 \\ \frac{a}N\dots\frac{a}N\end{matrix}\right]\brc{\textbf{z}\,|\,T}}
{\theta\brc{\textbf{z}\,|\,T}},\quad a=1\dots N-1,}
where we use the Riemann theta function\footnote{The Riemann theta function with
characteristics $\boldsymbol{a},\boldsymbol{b}\in\mathbb{Q}$  and $g\times g$ period matrix $T$ is
$$\theta\left[\begin{matrix}\boldsymbol{a}
\\ \boldsymbol{b}\end{matrix}\right](\boldsymbol{z}\,|\,T)=
\sum_{\boldsymbol{n}\in\mathbb{Z}^g}\mathrm{exp}
\brc{\imath\pi(\boldsymbol{n}+\boldsymbol{a})^t T(\boldsymbol{n}+\boldsymbol{a})
+2\imath\pi
(\boldsymbol{n}+\boldsymbol{a})\cdot(\boldsymbol{z}+\boldsymbol{b})
}
$$
where ${\bf a}$, ${\bf b}$ and ${\bf n}$ are $g$-dimensional vectors.} of genus $g=N$ with the $N\times N$ period matrix $T$ of the corresponding Seiberg-Witten curve. This matrix is given by the prepotential $\mathcal{F}$ and is a function of just $N$ flat moduli $a_i$:
\eq{T_{ij}=\frac{\partial^2\mathcal{F}}{\partial a_i\partial a_j}.}
In the previous works, the special property of the elliptic Calogero spectral curve was used to reduce the genus of theta functions in (\ref{eq:dellH}) to $N-1$. Here we introduce the genus $N$ theta functions to obtain the equations for the prepotential.
Also, the non-reduced form (\ref{eq:dellH}) allows one to study some other spectral curves, for example, corresponding to the theories with matter in the fundamental representation.

The Poisson commutativity of the Hamiltonians
\eqlb{eq:Pcomm}{\bfi{H_a,\, H_b}=0}
is considered with respect to the Seiberg-Witten symplectic structure
\eq{\omega^{\textrm{SW}}=\sum_{i=1}^{N}\rmd z_i\wedge\rmd a_i.}
Since (\ref{eq:Pcomm}) should be valid for arbitrary values of $\textbf{z}$, one can rewrite the involutivity conditions as a system of equations depending on the period matrix and its derivatives only. In \cite{ABMMZ}, this was done with the help of the standard basis in the linear space of weight $3$ theta functions:
\eq{\theta\left[\begin{matrix} \vec{\alpha}/3\\ (a+b)/N \ldots (a+b)/N \end{matrix} \right]
\brc{3\textbf{z}\,|\,3 T},\quad 0\leq\alpha_i<3.}
The result is the following set of $z$-independent relations equivalent to the commutativity conditions (\ref{eq:Pcomm}):
\eqlb{eq:invCon}{\boxed{
\forall\vec{\alpha}\in\mathbb{Z}^g/3\mathbb{Z}^g:\quad
\sum_{i,j,k=1}^{N}\frac{\partial^3 \mathcal{F}}{\partial a_i\partial a_j\partial a_k} C^{\vec{\alpha}}_{ijk}=0,}}
where
\eqn{C^{\vec{\alpha}}_{ijk}=
\sum_{\vec{\beta}\in\mathbb{Z}^g/2\mathbb{Z}^g}
\left(9\theta'_{i}
\left[ \begin{matrix} \frac{\vec{\beta}-\vec{\alpha}}2\\ a/N\ldots a/N \end{matrix} \right]
\brc{0\,|\,2 T}\,
\theta''_{jk}
	\bsq{\at{\frac{\vec{\beta}}2-\frac{\vec{\alpha}}6}{\brc{2\text{b}-\text{a}}/N \ldots\brc{2\text{b}-\text{a}}/N}}
\brc{0\,|\,6 T}-
\right.}
\eqlb{eq:thetacoord}{\left.-
\theta'''_{ijk}
\left[ \begin{matrix}\frac{\vec{\beta}-\vec{\alpha}}2\\ a/N \ldots a/N \end{matrix} \right]
\brc{0\,|\,2 T}\,
\theta\bsq{\at{\frac{\vec{\beta}}2-\frac{\vec{\alpha}}6}{\brc{2\text{b}-\text{a}}/N \ldots\brc{2\text{b}-\text{a}}/N}} \brc{0\,|\,6 T}\right).
}
The theta constants (\ref{eq:thetacoord}) have the following Fourier expansion:
\eqn{
C^{\vec{\alpha}}_{ijk}=4\,\brc{\pi\imath}^3
\exp\brc{-2\pi\imath\,\frac{(a+b)\sum_l{\alpha_l}}{3N}}\times}
\eqlb{eq:CFour}{\times\sum_{\textbf{m},\textbf{n}\in\mathbb{Z}^g}
\exp\brc{2\pi\imath\,\brc{\textbf{m}+\frac{\textbf{n}-\vec{\alpha}}2}T \brc{\textbf{m}+\frac{\textbf{n}-\vec{\alpha}}2}+
2\pi\imath\,
\brc{\frac{\textbf{n}}2-\frac{\vec{\alpha}}6}3T
\brc{\frac{\textbf{n}}2-\frac{\vec{\alpha}}6}}\times}
\eqn{\times\exp\brc{2\pi\imath\,\frac{\text{a}\sum_l m_l+\text{b}\sum_l n_l}N} \abs{\vec{\alpha}-\textbf{m}-\textbf{n},\,\textbf{m},\,\textbf{n}}_{ijk}}
with
\eq{\abs{\textbf{n},\textbf{m},\textbf{l}}_{ijk}=\left|\matr{ccc}
{1 & n_i & n_j n_k\\
 1 & m_i & m_j m_k\\
 1 & l_i & l_j l_k
}\right|.}

It was also proven in \cite{ABMMZ} that for $N=3$ relations (\ref{eq:invCon}) with \eqs{\exp\brc{\frac{2\pi\imath}3\sum_l\alpha_l}\neq 1} are trivial. The nontrivial relations can be reduced to the form
\eqlb{eq:inv3}{\exp
\brc{\frac{2\pi\imath}3\sum_{l=1}^{3}\alpha_l}=1:\quad
\sum_{i=1}^{3}C_i^{\vec{\alpha}}=0}
with
\eq{C_i^{\vec{\alpha}}=\sum_{\vec{\beta}\in\mathbb{Z}^3/2\mathbb{Z}^3}
\theta'_{i}
\left[ \begin{matrix} \frac{\vec{\beta}-\vec{\alpha}}2\\ 1/3\ldots 1/3 \end{matrix} \right]
\brc{0\,|\,2 T}\,
\frac{\partial}{\partial a_i}\theta
	\bsq{\at{\frac{\vec{\beta}}2-\frac{\vec{\alpha}}6}{1\ldots 1}}
\brc{0\,|\,6 T}.}

\section{Equations for the perturbative Seiberg-Witten prepotential}
\label{sec:pert_SW_prep}
The Seiberg-Witten prepotentials are usually presented as a sum of free parts: the classical, perturbative and instanton ones
\eq{\mathcal{F}=\mathcal{F}^{\textrm{class}}+ \mathcal{F}^{\textrm{pert}}+ \mathcal{F}^{\textrm{inst}}.}
We consider the special class of elliptic integrable systems associated with the low-energy limit of ${\cal N}=2$ SUSY gauge theories with adjoint matter hypermultiplet. In this class of systems, the following general expression for the prepotential holds:
\eqlb{eq:genPrep}{\mathcal{F}=\frac12\tau\sum_{i=1}^{N}a_i^2 + \frac12m \tau\sum_{i=1}^{N}a_i+\mathcal{F}^{\textrm{pert}}+\sum_{k\in\mathbb{N}}q^k\,\mathcal{F}^{(k)},}
where $m$ is the mass of the hypermultiplet, $q=\exp\brc{2\pi\imath\,\tau}$ and the elliptic parameter $\tau$  is related to the gauge coupling $e$ and to the $\theta$-angle of the gauge theory in the following way:
\eq{\tau= \frac{\theta}{2\pi}+\frac{4\pi\imath}{e^2}.}
Another important property of the systems under consideration is the following condition on the period matrix $T_{ij}$:
\eqlb{eq:ellT}{T_{ij}=\frac{\partial^2\mathcal{F}}{\partial a_i\partial a_j},\qq
\forall i:\quad\sum_{j=1}^{N}T_{ij}=\tau,}
which means that the perturbative part $\mathcal{F}^{\textrm{pert}}$ and the instanton corrections $\mathcal{F}^{(k)}$ depend only on the differences $\brc{a_i - a_j}$ of the flat moduli $a_i$ instead of the moduli themselves.

Now the involutivity conditions (\ref{eq:invCon}) can be considered as non-linear equations on the prepotential in form (\ref{eq:genPrep}). These equations depend on the second and the third partial derivatives of the prepotential with respect to the moduli $a_i$:
\eqlb{eq:pMT}{T_{ij}=\tau\, \delta_{ij} +
\frac{\partial^2\mathcal{F}^{\textrm{pert}}}{\partial a_i\partial a_j}
+\sum_{k\in\mathbb{N}}q^k\,
\frac{\partial^2\mathcal{F}^{(k)}}{\partial a_i\partial a_j},}
\eq{\frac{\partial^3 \mathcal{F}}{\partial a_i\partial a_j\partial a_k}=
\frac{\partial^3\mathcal{F}^{\textrm{pert}}}{\partial a_i\partial a_j\partial a_k}
+\sum_{k\in\mathbb{N}}q^k\,
\frac{\partial^3\mathcal{F}^{(k)}}{\partial a_i\partial a_j\partial a_k}.}
Since the theta constants $C^{\vec{\alpha}}_{ijk}$ from the involutivity conditions are exponentials of the period matrix (\ref{eq:pMT}), one gets a proper series expansion of (\ref{eq:invCon}) in powers of $q$. The equations on the perturbative prepotential  $\mathcal{F}^{\textrm{pert}}$ arise in the first non-zero order of this expansion.

In this section, we present equations for the perturbative Seiberg-Witten prepotential obtained from the involutivity conditions (\ref{eq:invCon}) with different vectors $\vec{\alpha}\in\mathbb{Z}^N/3\mathbb{Z}^N$. The expansion of $C^{\vec{\alpha}}_{ijk}$ in powers of $q$ can be derived with the help of the Fourier series (\ref{eq:CFour}).
For $\vec{\alpha}=\textbf{0}$, one has in the first non-zero order
\eqlb{eq:pertF}{\boxed{
\sum_{i=1}^{N}
\rme^{2\pi\imath\,\partial_i^2\mathcal{F}^{\textrm{pert}}}
\partial_i^3\mathcal{F}^{\textrm{pert}}=0,}}
where $\partial_i=\partial/\partial a_i$.
Next, we consider the equations corresponding to the vectors $\vec{\alpha}$ with two non-zero coordinates $\alpha_i$, $\alpha_j$, $i\neq j$:
\eqlb{eq:vec2}{\alpha_i=1,\quad \alpha_j=2,\quad \forall n\neq i,j:\quad \alpha_n=0.}
The first non-zero order in $q$ reads
\eqlb{eq:a2}{\boxed{
\forall i\neq j:\quad
\frac{\partial^3\mathcal{F}^{\textrm{pert}}}{\partial a_i^2\partial a_j}+
\frac{\partial^3\mathcal{F}^{\textrm{pert}}}{\partial a_i\partial a_j^2}=0.}}
The vectors $\vec{\alpha}$ with three non-zero coordinates $\alpha_i$, $\alpha_j$, $\alpha_k$,
$i\neq j\neq k$
\eqlb{eq:vec3}{\alpha_i=1,\quad \alpha_j=1,\quad \alpha_k=1,\quad \forall n\neq i,j,k:\quad \alpha_n=0}
give the following equations:
\eqn{\brc{\rme^{2\pi\imath\,\partial_{ij}^2\mathcal{F}^{\textrm{pert}}}+
\rme^{\partial_{ik}^2\mathcal{F}^{\textrm{pert}}}+
\rme^{\partial_{jk}^2\mathcal{F}^{\textrm{pert}}}
}\frac{\partial^3\mathcal{F}^{\textrm{pert}}}{\partial a_i\partial a_j\partial a_k}
+\rme^{2\pi\imath\,\partial_{ij}^2\mathcal{F}^{\textrm{pert}}}
\brc{\frac{\partial^3\mathcal{F}^{\textrm{pert}}}{\partial a_i^2\partial a_k}+
\frac{\partial^3\mathcal{F}^{\textrm{pert}}}{\partial a_j^2\partial a_k}}+}
\eqlb{eq:a3}{+\rme^{2\pi\imath\,\partial_{ik}^2\mathcal{F}^{\textrm{pert}}}
\brc{\frac{\partial^3\mathcal{F}^{\textrm{pert}}}{\partial a_i^2\partial a_j}+
\frac{\partial^3\mathcal{F}^{\textrm{pert}}}{\partial a_j\partial a_k^2}}
+\rme^{2\pi\imath\,\partial_{jk}^2\mathcal{F}^{\textrm{pert}}}
\brc{\frac{\partial^3\mathcal{F}^{\textrm{pert}}}{\partial a_i\partial a_j^2}+
\frac{\partial^3\mathcal{F}^{\textrm{pert}}}{\partial a_i\partial a_k^2}}
=0.}
Equations for other non-zero vectors $\vec{\alpha}$ different from (\ref{eq:vec2}) and (\ref{eq:vec3})  are more complicated, so we do not write them down explicitly.
However, the whole system of equations (\ref{eq:invCon}) along with condition (\ref{eq:ellT}) provides the following set of equations:
\eqlb{eq:Fijk}{\boxed{
\forall i\neq j\neq k:\quad
\frac{\partial^3\mathcal{F}^{\textrm{pert}}}{\partial a_i\partial a_j\partial a_k}=0,}}
which was proven for $N=3,4,5$.

Since the Seiberg-Witten prepotentials are invariant under any permutation of the flat moduli $a_i$, solutions to the equations (\ref{eq:a2}) and (\ref{eq:Fijk}) are described by the following class of perturbative prepotentials:
\eqlb{eq:Fpij}{i\neq j:\quad\frac{\partial^2}{\partial a_i\partial a_j}\mathcal{F}^{\textrm{pert}}=-\frac1{2\pi\imath}
\log\brc{1-\frac{m^2}{f\brc{a_i-a_j}}},
}
where $f\brc{x}$ is an even function. In the known cases of elliptic integrable systems, this function reduces to $x^2$ for the elliptic Calogero-Moser system, to $\sinh\brc{x}^2$ for the elliptic Ruijsenaars system and to $\sn\brc{x|\,\hat{\tau}}^2$ for the double-elliptic system.

Now, using equations (\ref{eq:pertF}) and (\ref{eq:a3}), one can define the most general form of the function $f\brc{x}$. Here we would like to point out that our considerations are restricted by the strong condition, (\ref{eq:ellT}), which gives
\eq{\frac{\partial^2}{\partial a_i^2}\mathcal{F}^{\textrm{pert}}=-
\sum_{\substack{j=1 \\ j\neq i}}^{N}
\frac{\partial^2}{\partial a_i\partial a_j}\mathcal{F}^{\textrm{pert}}.}
If one drops the condition (\ref{eq:ellT}), the diagonal elements of the period matrix become independent of the non-diagonal ones and equation (\ref{eq:pertF}) turns out to be essentially different from equation (\ref{eq:a3}). This case could correspond to the theories with matter in the fundamental representation.

Introducing the notation
\eqlb{eq:F0ij}{F^{(0)}_{ij}=\rme^{-2\pi\imath\,\partial_{ij}^2\mathcal{F}^{\textrm{pert}}}=
1-\frac{m^2}{f\brc{a_i-a_j}},\quad
F^{(0)}_{ij|k}=\frac{\partial}{\partial a_k}\brc{1-\frac{m^2}{f\brc{a_i-a_j}}},\quad
i\neq j,}
we rewrite equation (\ref{eq:pertF}) in the rational form. For the first non-trivial cases of
three and four particles, one has
\eqlb{eq:F03}{N=3:\quad F^{(0)}_{13}\, F^{(0)}_{12|1}+F^{(0)}_{23}\, F^{(0)}_{12|2}+F^{(0)}_{12}\, F^{(0)}_{13|1}+
F^{(0)}_{23}\, F^{(0)}_{13|3}+F^{(0)}_{12}\, F^{(0)}_{23|2}+ F^{(0)}_{13}\, F^{(0)}_{23|3}=0,}
\eqn{N=4:\quad F^{(0)}_{13} F^{(0)}_{14} F^{(0)}_{12|1}+F^{(0)}_{23} F^{(0)}_{24} F^{(0)}_{12|2}+F^{(0)}_{12} F^{(0)}_{14} F^{(0)}_{13|1}+F^{(0)}_{23} F^{(0)}_{34} F^{(0)}_{13|3}+F^{(0)}_{12} F^{(0)}_{13} F^{(0)}_{14|1}+F^{(0)}_{24} F^{(0)}_{34} F^{(0)}_{14|4}+}
\eqlb{eq:F04}{+F^{(0)}_{12} F^{(0)}_{24} F^{(0)}_{23|2}+F^{(0)}_{13} F^{(0)}_{34} F^{(0)}_{23|3}+F^{(0)}_{12} F^{(0)}_{23} F^{(0)}_{24|2}+F^{(0)}_{14} F^{(0)}_{34} F^{(0)}_{24|4}+F^{(0)}_{13} F^{(0)}_{23} F^{(0)}_{34|3}+F^{(0)}_{14} F^{(0)}_{24} F^{(0)}_{34|4}=0.}
Consider the series expansion for an even function $f\brc{x}$
\eq{f\brc{x}=\sum_{n\in\mathbb{Z}}^{\infty}\hat{e}_{n-1}\,x^{2n}.}
Besides the trivial solution $f\brc{x}=const$, the both equations, (\ref{eq:F03}) and (\ref{eq:F04}) admit the following series solution:
\eqlb{eq:fexp}{f\brc{x}=x^2 + \sum_{n=2}^{\infty}e_{n-1}\,x^{2n},}
where the coefficient $e_0$ is rescaled with the help of mass parameter $m$ in (\ref{eq:F0ij}).
Substituting (\ref{eq:fexp}) in (\ref{eq:F03}) and (\ref{eq:F04}) and solving each equation with respect to the coefficients $e_n$, one gets the recurrence relations
\eq{e_4=\frac{2}{3}e_1^4 - \frac{7}{3} e_1^2\, e_2 +2 e_1\, e_3+ \frac{2}{3} e_2^2,}
\eq{e_5= \frac{20}{33}e_1^5-\frac{49}{33} e_1^3\, e_2 + \frac{14}{11} e_1^2\, e_3 - \frac{37}{33} e_1\, e_2^2+\frac{19}{11} e_2\, e_3,}
\eq{\ldots}
and so on (see Appendix \ref{sec:recf}).
The same formulae are valid for equation (\ref{eq:a3}).
The most general function satisfying the recurrence relations (\ref{eq:e4B})--(\ref{eq:e10B}) is
\eq{f\brc{x}=\frac{\sn\brc{\beta\,x|\,\hat{\tau}}^2}
{\beta^2 -\gamma\,\sn\brc{\beta\,x|\,\hat{\tau}}^2},\quad
1-\frac{m^2}{f\brc{x}}=1+m^2\,\gamma-
\frac{m^2\,\beta^2}{\sn\brc{\beta\,x|\,\hat{\tau}}^2},}
where the first parameter $\beta^{-1}$ corresponds to the first period $\hat{\omega}_1$ of another, second torus with the elliptic parameter $\hat{\tau}=\hat{\omega}_2/\hat{\omega}_1$. The second parameter $\gamma$ corresponds to the simple shift in the classical prepotential $\mathcal{F}^{\textrm{class}}$ and the rescaling of the mass $m$, which can be seen from expression (\ref{eq:Fpij}) for the perturbative prepotential. As a result, there is one essential parameter $\hat{\tau}$ corresponding to the elliptic parameter of the second torus in the double-elliptic system. Finally, we present the most general solution (with respect to the second partial derivatives) to equations (\ref{eq:pertF}), (\ref{eq:a2}), (\ref{eq:a3})  and (\ref{eq:Fijk}) with property (\ref{eq:ellT}):
\eqlb{eq:pFsol}{i\neq j:\quad\frac{\partial^2}{\partial a_i\partial a_j}\mathcal{F}^{\textrm{pert}}=-\frac1{2\pi\imath}
\log\brc{1-\frac{m^2}{\sn\brc{\beta\brc{a_i-a_j}|\,\hat{\tau}}^2}},\quad
\frac{\partial^2}{\partial a_i^2}\mathcal{F}^{\textrm{pert}}=-
\sum_{\substack{j=1 \\ j\neq i}}^{N}
\frac{\partial^2}{\partial a_i\partial a_j}\mathcal{F}^{\textrm{pert}}.
}

\section{Non-perturbative corrections for $N=3$}
\label{sec:Inst3}
In this section, we describe the method of constructing non-perturbative solutions of (\ref{eq:invCon}), which is based on the series expansion in powers of $q$. As we mentioned earlier, the first non-zero order in $q$ depends only on the perturbative part of the prepotential; the second non-zero order incorporates the perturbative part and the first instanton correction and so on.

We start with the leading, perturbative order of (\ref{eq:invCon}). For $N=3$, there are $5$ different equations corresponding to different vectors
$\vec{\alpha}\in\mathbb{Z}^3/3\mathbb{Z}^3$:
\eqlb{eq:p000}{\vec{\alpha}=\brc{0,0,0}:\quad \sum_{i=1}^{3}
\rme^{2\pi\imath\,\partial_i^2\mathcal{F}^{\textrm{pert}}}
\partial_i^3\mathcal{F}^{\textrm{pert}}=0,}
\eqlb{eq:p012}{\vec{\alpha}=\brc{0,1,2}:\quad
\frac{\partial^3\mathcal{F}^{\textrm{pert}}}{\partial a_2^2\partial a_3}+
\frac{\partial^3\mathcal{F}^{\textrm{pert}}}{\partial a_2\partial a_3^2}=0,}
\eqlb{eq:p102}{\vec{\alpha}=\brc{1,0,2}:\quad
\frac{\partial^3\mathcal{F}^{\textrm{pert}}}{\partial a_1^2\partial a_3}+
\frac{\partial^3\mathcal{F}^{\textrm{pert}}}{\partial a_1\partial a_3^2}=0,}
\eqlb{eq:p120}{\vec{\alpha}=\brc{1,2,0}:\quad
\frac{\partial^3\mathcal{F}^{\textrm{pert}}}{\partial a_1^2\partial a_2}+
\frac{\partial^3\mathcal{F}^{\textrm{pert}}}{\partial a_1\partial a_2^2}=0,}
\eqlb{eq:p111}{\vec{\alpha}=\brc{1,1,1}:\quad
\frac{\partial^3\mathcal{F}^{\textrm{pert}}}{\partial a_1\partial a_2\partial a_3}=0,}
where the last equation was simplified with the help of (\ref{eq:ellT}).

Expanding (\ref{eq:invCon}) up to the second non-zero order in $q$, one gets non-perturbative corrections to the above equations. The resulting equations depend on the perturbative part of the prepotential $\mathcal{F}^{\textrm{pert}}$ and on the first instanton correction $\mathcal{F}^{(1)}$. We use the relations obtained in the perturbative order and the notation $F^{(0)}_{ij}$ from (\ref{eq:F0ij}) to simplify the first non-perturbative corrections to (\ref{eq:p000})--(\ref{eq:p111}), which acquire the form
\eqn{2\brc{F^{(0)}_{12}-F^{(0)}_{23}}\brc{F^{(0)}_{13}-F^{(0)}_{23}}
\brc{F^{(0)}_{12}+F^{(0)}_{13}+F^{(0)}_{23}}
\rme^{2\pi\imath\,\partial_1^2\mathcal{F}^{\textrm{pert}}}
\partial_1^3\mathcal{F}^{\textrm{pert}}+}
\eqlb{eq:a0F1}{+
\sum_{i=1}^{3}\rme^{2\pi\imath\,
\partial_i^2\mathcal{F}^{\textrm{pert}}}
\brc{\partial_i^3\mathcal{F}^{(1)}
+2\pi\imath\,
\partial_i^2\mathcal{F}^{(1)}
\partial_i^3\mathcal{F}^{\textrm{pert}}}=0,}
\eq{\frac{\partial^3\mathcal{F}^{(1)}}{\partial a_2^2\partial a_3}+
\frac{\partial^3\mathcal{F}^{(1)}}{\partial a_2\partial a_3^2}
+6\brc{F^{(0)}_{12}-F^{(0)}_{23}}\brc{F^{(0)}_{13}-F^{(0)}_{23}}
\brc{F^{(0)}_{12}
\frac{\partial^3\mathcal{F}^{\textrm{pert}}}{\partial a_1^2\partial a_2}
+F^{(0)}_{13}
\frac{\partial^3\mathcal{F}^{\textrm{pert}}}{\partial a_1^2\partial a_3}}=0,}
\eq{\frac{\partial^3\mathcal{F}^{(1)}}{\partial a_1^2\partial a_3}+
\frac{\partial^3\mathcal{F}^{(1)}}{\partial a_1\partial a_3^2}
+6\brc{F^{(0)}_{12}-F^{(0)}_{23}}\brc{F^{(0)}_{13}-F^{(0)}_{23}}
\brc{F^{(0)}_{12}
\frac{\partial^3\mathcal{F}^{\textrm{pert}}}{\partial a_1^2\partial a_2}
+F^{(0)}_{13}
\frac{\partial^3\mathcal{F}^{\textrm{pert}}}{\partial a_1^2\partial a_3}}=0,}
\eq{\frac{\partial^3\mathcal{F}^{(1)}}{\partial a_1^2\partial a_2}+
\frac{\partial^3\mathcal{F}^{(1)}}{\partial a_1\partial a_2^2}
+6\brc{F^{(0)}_{12}-F^{(0)}_{23}}\brc{F^{(0)}_{13}-F^{(0)}_{23}}
\brc{F^{(0)}_{12}
\frac{\partial^3\mathcal{F}^{\textrm{pert}}}{\partial a_1^2\partial a_2}
+F^{(0)}_{13}
\frac{\partial^3\mathcal{F}^{\textrm{pert}}}{\partial a_1^2\partial a_3}}=0,}
\eqlb{eq:a111F1}{\frac{\partial^3\mathcal{F}^{(1)}}{\partial a_1\partial a_2\partial a_3}-
6\brc{F^{(0)}_{12}-F^{(0)}_{23}}\brc{F^{(0)}_{13}-F^{(0)}_{23}}
\brc{F^{(0)}_{12}
\frac{\partial^3\mathcal{F}^{\textrm{pert}}}{\partial a_1^2\partial a_2}
+F^{(0)}_{13}
\frac{\partial^3\mathcal{F}^{\textrm{pert}}}{\partial a_1^2\partial a_3}}=0.}
One can calculate further non-perturbative equations up to any given order in $q$.
In each consequent order, new instanton corrections $\mathcal{F}^{\brc{k}}$ arise.
We performed these calculations up to the $8$-th non-zero order in the expansion of (\ref{eq:invCon}),
which provides equations for the first $7$ instanton corrections.

Now we describe some simple methods to obtain solutions of (\ref{eq:invCon}). Having the general solution (\ref{eq:pFsol}) for the perturbative prepotential, we substitute it into equations (\ref{eq:a0F1})--(\ref{eq:a111F1}) for the first instanton correction. First of all, these equations define the pole structure of the correction $\mathcal{F}^{(1)}$. Since functions $F^{(0)}_{ij}$ and the third derivatives of the perturbative prepotential $\mathcal{F}^{\textrm{pert}}$ exhibit poles only at $a_i=a_j$, $i\neq j$, the same is true for the first instanton correction. We note that the orders of poles are also restricted by the equations. The pole structure of the higher instanton corrections is defined in the same way by the preceding instanton corrections and the perturbative prepotential. Another important property is that the instanton part of the prepotential is a symmetric function of the differences $\brc{a_i-a_j}$. This property can be derived from equations (\ref{eq:invCon}) and the condition (\ref{eq:ellT}), we consider it as a natural ansatz.

To make use of these general considerations, we start with the elliptic Calogero-Moser system. In this case, the prepotential $\mathcal{F}^{\textrm{CM}}$ depends on two parameters: the elliptic modular parameter $\tau$ and the mass $m$. The first parameter $\tau$ gives the instanton expansion $\sum_k q^k \mathcal{F}^{\brc{k}}$. Each instanton correction $\mathcal{F}^{\brc{k}}$ can be further decomposed as a series into powers of the second parameter $m$.
This latter decomposition is also specified by equations (\ref{eq:invCon}) in each non-zero order in $q$. To make the calculations simpler, we use the homogeneity relation
\eqlb{eq:hom}{\sum_{i=1}^{N}a_i\frac{\partial\mathcal{F}^{\textrm{CM}}}{\partial a_i}+
m \frac{\partial\mathcal{F}^{\textrm{CM}}}{\partial m}-
2\mathcal{F}^{\textrm{CM}}=0.}
The calculations begin with equations (\ref{eq:a0F1})--(\ref{eq:a111F1}), where we use the perturbative prepotential corresponding to the elliptic Calogero-Moser system. Decomposing this equations into powers of $m$ and taking into the account the orders of poles at the points $a_i=a_j$, $i\neq j$, one gets a finite number of terms that could enter in the first instanton correction. In each term of a given degree of $m$ (and a given pole structure), the dependence on the flat moduli $a_i$ is fixed by the homogeneity relation (\ref{eq:hom}) and by the symmetric properties of the prepotential. Introducing a linear combination of these terms with undetermined coefficients $c_{\dots}$, we rewrite (\ref{eq:a0F1})--(\ref{eq:a111F1}) as a system of linear equations on the coefficients. Moving on to the next instanton corrections, we apply the same method of undetermined coefficients. Similar methods can be applied for the elliptic Ruijsenaars system and the double-elliptic system, which we discuss in sections \ref{sec:RS} and \ref{sec:Dell}.

\section{Elliptic Calogero-Moser system and $4d$ prepotential}
\label{sec:InstCM}
In the case of the elliptic Calogero-Moser system, the first two instanton corrections were computed in \cite{HPh'98} with the help of the spectral curve
\eqlb{eq:CMcurve}{\det\brc{L\brc{z}-k\,I}=0,}
of the Seiberg-Witten differential $\rmd\lambda = k\,\rmd z$
and of a renormalization group equation for the variation of $\mathcal{F}$ with respect to $\tau$ \cite{HPh'98}:
\eqlb{eq:prepRG}{\frac{\partial\mathcal{F}^{\textrm{CM}}}{\partial\tau}=\frac{2\omega_1}{8\pi^2}\sum_{j=1}^{N}\oint_{A_j} k^2\rmd z,}
where the right hand side coincides with the second order Hamiltonian of the Calogero-Moser system \cite{Calogero'75,Calogero'76,Moser'75,HPh'98} up to some $a_i$-independent term:
\eqlb{eq:FECMH}{\frac{\partial\mathcal{F}^{\textrm{CM}}}{\partial\tau}=-\frac{\omega_1^2}{2\pi^2}
\brc{\sum_{i=1}^{N}p_i^2-m^2\sum_{i<j}\wp\brc{u_i-u_j}}+const.}
We compute first $4$ instanton corrections using the curve (\ref{eq:CMcurve}) in Appendix \ref{sec:ECMcurve}.
In this section, we use equations (\ref{eq:invCon}) to define the structure of the instanton part of the $4d$ prepotential.

\subsection{Instanton expansion}
For the elliptic Calogero-Moser system, the perturbative part of the prepotential is
\eqlb{eq:FCMpert}{\mathcal{F}^{\textrm{CM,pert}}=\frac{1}{8\pi\imath}
\sum_{i,j=1}^{N}\brc{\brc{a_i-a_j+m}^2\log\brc{a_i-a_j+m}^2-
\brc{a_i-a_j}^2\log\brc{a_i-a_j}^2}.}
Using this expression and equations (\ref{eq:invCon}), one can calculate the instanton corrections to the prepotential $\mathcal{F}^{\textrm{CM}}$ as it was described in section \ref{sec:Inst3}. Introducing the new variables
\eq{
\begin{array}{c}
\label{eq:svar_CM}
s_{ij}\brc{\textbf{a}}=\frac14\left(\brc{a_{12}}^{2i}\brc{a_{13}}^{2j}+
\brc{a_{12}}^{2i}\brc{a_{23}}^{2j} +\brc{a_{13}}^{2i}\brc{a_{23}}^{2j}+
\brc{i\leftrightarrow j}
\right),\\
t\brc{\textbf{a}}=\brc{a_{12}}^2\brc{a_{13}}^2 \brc{a_{23}}^2\\
\end{array}
}
with the notation \eqs{a_{ij}=\brc{a_i-a_j}}, we get the following expansion for the instanton part of the prepotential:
\eqlb{eq:prepCM}{\boxed{
\mathcal{F}^{\textrm{CM,inst}}=
\frac{m^2}{\pi\imath}\sum_{n\in\mathbb{N}}\,\sum_{i=0}^{1}\,\sum_{j=i}^{n}
\,m^{6n-2i n-2j}\,c_{n,i n,j}\brc{\tau}\,\frac{s_{in,j}}{t^{n}},}}
where
\eq{c_{n,i,j}\brc{\tau}=\sum_{l\geq(n+1)/2} c_{n,i,j,l}\,q^l}
and the coefficients $c_{n,i,j,l}$ are rational.

Equations (\ref{eq:invCon}) allow one to compute the coefficients $c_{n,i,j,l}$ up to any finite instanton order. We computed the first $7$ instanton corrections and the results suggest that
the functions $c_{n,i,j}\brc{\tau}$ are quasimodular forms of level $1$ and of weight $6n-2i-2j$ up to some constant shifts (coming from the perturbative part of the prepotential):
\eqlb{eq:c1_CM}{c_{111}=\frac1{12}-\frac{E_2}{12},\qq
c_{101}=\frac1{288}\brc{E_2^2-E_4},\qq
c_{222}=\frac1{60}-\frac1{360}\brc{5\,E_2^2+E_4}}
and so on. This fact is in a perfect agreement with the modular properties of the curve (\ref{eq:CMcurve}), which we discuss below in section \ref{subsec:Mod_CM}.

Since the quasimodular forms of level $1$ form a polynomial ring over the complex numbers in three generators (the Eisenstein series), \cite{Kob}:
\eq{E_2\brc{\tau}=1-24\sum _{n\in\mathbb{N}}\frac{n\,q^n}{1-q^n},}
\eq{E_4\brc{\tau}=1+240\sum _{n\in\mathbb{N}}\frac{n^3\,q^n}{1-q^n},}
\eq{E_6\brc{\tau}=1-504\sum _{n\in\mathbb{N}}\frac{n^5\,q^n}{1-q^n},}
computing the coefficients $c_{n,i,j,l}$ up to any finite instanton order allows one to obtain exact expressions for the functions \eqs{c_{n,i,j}\brc{\tau}} with small enough weights.
In particular, the first $7$ instanton corrections allow one to determine the functions \eqs{c_{n,i,j}\brc{\tau}} up to weight $14$:
\eqlb{eq:cTab}{
\begin{array}{c c}
\textrm{Weight} & \textrm{Functions}\\
2 & c_{111} \\
4 & c_{101}\quad c_{222}\\
6 & c_{100}\quad c_{221}\quad c_{333}\\
8 & c_{202}\quad c_{332}\quad c_{444}\\
10 & c_{201}\quad c_{331}\quad c_{443}\quad c_{555}\\
12 & c_{200}\quad c_{303}\quad c_{442}\quad c_{554}\quad c_{666}\\
14 & c_{302}\quad c_{441}\quad c_{553}\quad c_{665}\quad c_{777}\\
\end{array}}
and the results are presented in Appendix \ref{sec:4d_c}.

\subsection{Modular properties}
\label{subsec:Mod_CM}
The spectral curve of the elliptic Calogero-Moser system (\ref{eq:CMcurve}) is given by the Lax matrix
\eq{L_{ij}=p_i\delta_{ij}-m\brc{1-\delta_{ij}}\frac{\sigma\brc{z-u_i+u_j}}{\sigma\brc{z}\sigma\brc{u_i-u_j}},}
\eq{\sigma\brc{z}=\prod_{n_1,n_2}\brc{1-\frac{z}{n_1\omega_1+n_2\omega_2}}
\exp\brc{\frac{z}{n_1\omega_1+n_2\omega_2}+\frac12\brc{\frac{z}{n_1\omega_1+n_2\omega_2}}^2}.}
The curve is invariant under the modular transformations
\eqlb{eq:modT}{\tau\rightarrow-\frac1{\tau}\quad \textrm{and}\quad \tau\rightarrow\tau+1}
of the elliptic parameter $\tau=\omega_2/\omega_1$.

Consider the definitions of the flat moduli $a_i$ and their duals $a^D_i$:
\eqlb{eq:a_Def}{a_i=\frac1{2\pi\imath}\oint_{A_i} k\rmd z,\quad
a^D_i=\frac1{2\pi\imath}\oint_{B_i} k\rmd z.}
Since under the transformation $\tau\rightarrow-1/\tau$ the cycles $A_i$ and $B_i$ interchange:
\eq{A_i \xrightarrow{\tau\rightarrow-1/\tau} B_i,\quad
B_i \xrightarrow{\tau\rightarrow-1/\tau}- A_i,}
the same do the moduli
\eqlb{eq:modTai}{a_i \xrightarrow{\tau\rightarrow-1/\tau} a^D_i,\quad
a^D_i \xrightarrow{\tau\rightarrow-1/\tau}- a_i.}
Thus, the period matrix $T^{\textrm{CM}}$
\eq{T^{\textrm{CM}}_{ij}=\frac{\partial a^D_i}{\partial a_j}=
\frac{\partial^2 \mathcal{F}^{\textrm{CM}}}{\partial a_i \partial a_j}}
transforms as
\eq{T^{\textrm{CM}} \xrightarrow{\tau\rightarrow-1/\tau} -\brc{T^{\textrm{CM}}}^{-1}.}
In other words, the following equation holds:
\eqlb{eq:Mod_CM}{T^{\textrm{CM}}\brc{\textbf{a}^D,\,m,\,-\tau^{-1}}=
-\brc{T^{\textrm{CM}}\brc{\textbf{a},\,m,\,\tau}}^{-1}.}
The second modular transformation from (\ref{eq:modT}) gives
\eq{T^{\textrm{CM}}_{ij}\brc{\textbf{a},\,m,\,\tau+1}=
T^{\textrm{CM}}_{ij}\brc{\textbf{a},\,m,\,\tau}+\delta_{ij}.}
The modular properties considered above were used by J.A. Minahan, D. Nemeschansky and N.P. Warner (MNW) \cite{MNW} to derive the modular anomaly equation:
\eqlb{eq:Mod_Eq_CM}{\frac{\partial \mathcal{F}^{\textrm{CM}}}{\partial E_2}=-\frac{\pi\imath}{12}
\sum_{i=1}^{N}\brc{\frac{\partial \mathcal{F}^{\textrm{CM}}}{\partial a_i}-\tau\,a_i}^2.}

However, equations (\ref{eq:Mod_CM}) and (\ref{eq:Mod_Eq_CM}) only describe dependence of the functions $c_{n,i,j}\brc{\tau}$ from (\ref{eq:prepCM}) on the quasimodular form $E_2$.
To obtain the exact expressions like (\ref{eq:c1_CM}), one needs to use some additional information.

\section{Elliptic Ruijsenaars system and $5d$ prepotential}
\label{sec:RS}
The spectral curve of the elliptic Ruijsenaars system \cite{BMMM'99} can be written in the form
\eqlb{eq:RScurve}{\sum_{n=0}^{\infty}\frac1{n!}\brc{\frac{m}{2\pi\imath}}^n
\partial_z^n\theta\bsq{\at{1/2}{1/2}}\brc{\pi z|\,\tau}\partial_k^n H\brc{k}=0,}
where
\eq{H\brc{k}=\prod_{i=1}^{N}\sinh\brc{\frac{\beta}2\brc{k-k_i}}.}
According to \cite{BMMM'99,Nek5}, this curve corresponds to the five-dimensional theory. In the limit $\beta\rightarrow0$, one gets the spectral curve of the elliptic Calogero-Moser system with $H\brc{k}=\prod_{i=1}^{N}\brc{k-k_i}$.
In principle, the curve (\ref{eq:RScurve}) could be used to calculate the $5d$ prepotential, as it was done \cite{HPh'98} in the case of the $4d$ prepotential and the spectral curve (\ref{eq:CMcurve}).
The resulting prepotential would depend on the three parameters $\tau$, $m$ and $\beta$.

\subsection{Instanton expansion}
As earlier, for finding the non-perturbative corrections to the prepotential, we start with the perturbative part
\eqlb{eq:pertRS}{i\neq j:\quad\frac{\partial^2}{\partial a_i\partial a_j}\mathcal{F}^{\textrm{RS,pert}}=-\frac1{2\pi\imath}
\log\brc{1-\frac{m^2}{\sinh\brc{\beta\brc{a_i-a_j}}^2}},\quad
\frac{\partial^2}{\partial a_i^2}\mathcal{F}^{\textrm{RS,pert}}=-
\sum_{\substack{j=1 \\ j\neq i}}^{N}
\frac{\partial^2}{\partial a_i\partial a_j}\mathcal{F}^{\textrm{RS,pert}}}
and use equations (\ref{eq:a0F1})--(\ref{eq:a111F1}) to define the first instanton correction. These equations can be decomposed in powers of $m$ (with a finite number of terms in the decomposition) and solved order by order.
After solving the equations for some first orders in $m$, we introduce the following ansatz: the instanton part of the $5d$ prepotential is a symmetric function of $\bfi{\sinh\brc{\beta\brc{a_i-a_j}}}$. This allows one to use the same method of undetermined coefficients, as in the $4d$ case. The only difference is that the homogeneity relation (\ref{eq:hom}) does not work for the $5d$ prepotential and the undetermined coefficients $c_{\dots}$ acquire the series expansion in powers of $m$. We probed this ansatz in the first $7$ instanton corrections and calculated the corresponding coefficients.

To present the results, we introduce the following functions:
\eq{
\label{eq:svar_RS}
\begin{array}{c}
\tilde s_{ij}\brc{\textbf{a},\,\beta} =\frac14\left(\sinh\brc{\beta\,a_{12}}^{2i}\sinh\brc{\beta\,a_{13}}^{2j}+
\sinh\brc{\beta\,a_{12}}^{2i}\sinh\brc{\beta\,a_{23}}^{2j}+\right.\\
\left. +\sinh\brc{\beta\,a_{13}}^{2i}\sinh\brc{\beta\,a_{23}}^{2j}+
\brc{i\leftrightarrow j}\right),\\
\tilde t\brc{\textbf{a},\,\beta}=\sinh\brc{\beta\,a_{12}}^2\sinh\brc{\beta\,a_{13}}^2
\sinh\brc{\beta\,a_{23}}^2.\\
\end{array}}
Then, the instanton part of the prepotential $\mathcal{F}^{\textrm{RS}}$ for $N=3$ acquires the form
\eqlb{eq:prepRS}{\boxed{
\mathcal{F}^{\textrm{RS,inst}}=
\frac{m^2}{\pi\imath\,\beta^2}\sum_{n\in\mathbb{N}}\,\sum_{i=0}^{1}\,
\sum_{j=i}^{n}\,m^{6n-2in-2j}\,c_{n,in,j}\brc{m,\tau}\,
\frac{\tilde s_{in,j}}{\tilde t^{n}},}}
where the functions $c_{n,i,j}\brc{m,\tau}$ admit the series expansion
\eqlb{eq:RS_c}{c_{n,i,j}\brc{m,\tau}=\sum_{l\geq\brc{n+1}/2}\sum_{k=0}^{2l-n-1}\,m^{2k}\, c_{n,i,j,k,l}\,q^l}
with rational coefficients $c_{n,i,j,k,l}$. The summation over indices $i$ and $j$ in (\ref{eq:prepRS}) is taken specifically to avoid uncertainties related to the identities like
\eq{s_{02}-2\,s_{11}=2\,t,}
\eq{s_{03}-2\,s_{12}-4\,t\,s_{01}=3\, t,}
\eq{2\,s_{22}-2\,s_{13}+4\,t\,s_{11}+3\,t\,s_{01}=0}
and similar ones for other functions $s_{ij}$.

The coefficients in the expansions (\ref{eq:RS_c}) are connected with the ring of quasimodular forms of level $1$ in the following way.
Consider functions of the elliptic parameter $\tau$
\eq{c_{n,i,j,k}\brc{\tau}=\sum_{l\geq\brc{n+1}/2} c_{n,i,j,k,l}\,q^l,}
so that
\eq{c_{n,i,j}\brc{m,\tau}=\sum_{k=0}^{+\infty}\,m^{2k}\, c_{n,i,j,k}\brc{\tau}.}
Then $c_{n,i,j,0}\brc{\tau}$ coincide with the $4d$ functions $c_{n,i,j}\brc{\tau}$:
\eq{c^{\brc{5\textrm{d}}}_{n,i,j,0}\brc{\tau}=c^{\brc{4\textrm{d}}}_{n,i,j}\brc{\tau}.}
Other functions $c_{n,i,j,k}\brc{\tau}$ with $k>0$ are linear combinations of the quasimodular forms with different weights that are not greater than $6n-2i-2j+2k$ (up to some constant shifts):
\eq{c_{1111}=-\frac2{45}+\frac1{18}E_2+\frac1{360}\brc{-5\,E_2+E_4},}
\eq{c_{1011}=-\frac1{288}\brc{E_2^2-E_4}+\frac1{12\,960}\brc{25\,E_2^3-33\,E_2\,E_4+8\,E_6},}
\eq{c_{2221}=-\frac3{280}+\frac1{360}\brc{5\,E_2^2+E_4}-
\frac1{45\,360}\brc{245\,E_2^3+42\,E_2\,E_4-17\,E_6}}
and so on.

The peculiar properties of $c_{n,i,j,k}\brc{\tau}$ described above are due to the non-canonical choice of the parameters in the $5d$ prepotential. The parameter $m$ is natural for the non-linear equations under consideration, since in each finite order in $q$ the equations have finite expansions in powers of $m$. In Seiberg-Witten theory \cite{BMMM1'99}, the natural choice of the parameters is different: the three parameters are $\tau$, $\beta$ and $\epsilon$. Comparing the second partial derivative of the perturbative $5d$ prepotential in the form (\ref{eq:svar_RS}) and the results from \cite{BMMM1'99}, we establish the connection between the parameters $m$ and $\epsilon$ as
\eq{m=\sinh\brc{\epsilon}.}
Now, one can rewrite the whole $5d$ prepotential as a series in $\epsilon$:
\eqlb{eq:prepRS_full}{\boxed{
\mathcal{F}^{\textrm{RS}}=\frac12\tau\sum_{i=1}^{3}a_i^2 +
\frac{\epsilon^2}{4\pi\imath\,\beta^2} \sum_{i<j}\log\brc{\sinh\brc{\beta\,a_{ij}}^2}
+
\frac{\epsilon^2}{\pi\imath\,\beta^2}\sum_{n\in\mathbb{N}}\,\sum_{i=0}^{1}\,
\sum_{j=i}^{n}\,\epsilon^{6n-2in-2j}\,\tilde {c}_{n,in,j}\brc{\epsilon,\tau}\,
\frac{\tilde s_{in,j}}{\tilde t^{n}},}}
where
\eq{\tilde {c}_{n,i,j}\brc{\epsilon,\tau}=\sum_{k=0}^{+\infty}\,\epsilon^{2k}\, \tilde {c}_{n,i,j,k}\brc{\tau}}
and the functions $\tilde {c}_{n,i,j,k}\brc{\tau}$ are proper quasimodular forms of weight $6n-2i-2j+2k$:
\eq{\tilde {c}_{1111}=\frac1{360}\brc{-5\,E^2_2+E_4},}
\eq{\tilde {c}_{1011}=\frac1{12\,960}\brc{25\,E_2^3-33\,E_2\,E_4+8\,E_6},}
\eq{\tilde {c}_{2221}=-
\frac1{45\,360}\brc{245\,E_2^3+42\,E_2\,E_4-17\,E_6}.}
Using the same method of obtaining exact expressions as in the $4d$ case,
we determine the functions $\tilde {c}_{n,i,j,k}\brc{\tau}$, $k>0$ up to weight $14$ in Appendix \ref{sec:5d_c}.

\subsection{Modular properties}
The spectral curve of the elliptic Ruijsenaars system \cite{BMMM'99} is invariant under the transformations
\eq{\tau\rightarrow-\frac1{\tau},\quad \epsilon\rightarrow \frac{\epsilon}{\tau},\quad
\beta\rightarrow \frac{\beta}{\tau} \quad \textrm{and}\quad \tau\rightarrow\tau+1.}

The definitions of the flat moduli $a_i$ and their duals $a^D_i$ are exactly the same as in the $4d$ case (\ref{eq:a_Def}). This provides us with the following equations for the period matrix $T^{\textrm{RS}}$:
\eqlb{eq:Mod_RS}{T^{\textrm{RS}}\brc{\textbf{a}^D,\,\frac{\epsilon}{\tau},\,\frac{\beta}{\tau},
\,-\frac1{\tau}}=-\brc{T^{\textrm{RS}}\brc{\textbf{a},\,\epsilon,\,\beta,\,\tau}}^{-1},}
\eq{T^{\textrm{RS}}_{ij}\brc{\textbf{a},\,\epsilon,\,\beta,\,\tau+1}=
T^{\textrm{RS}}_{ij}\brc{\textbf{a},\,\epsilon,\,\beta,\,\tau}+\delta_{ij}}
and the modular anomaly equation in the MNW form:
\eqlb{eq:Eq_Mod_RS}{\frac{\partial \mathcal{F}^{\textrm{RS}}}{\partial E_2}=-\frac{\pi\imath}{12}
\sum_{i=1}^{N}\brc{\frac{\partial \mathcal{F}^{\textrm{RS}}}{\partial a_i}-\tau\,a_i}^2.}
Equations (\ref{eq:Mod_RS}) and (\ref{eq:Eq_Mod_RS}) describe dependence of the functions $\tilde {c}_{n,i,j,k}\brc{\tau}$ from (\ref{eq:prepRS_full}) on the quasimodular form $E_2$.

\section{Double-elliptic system and $6d$ prepotential}
\label{sec:Dell}

The double-elliptic system corresponds to the most general solution of equations (\ref{eq:invCon}) with the property (\ref{eq:ellT}). As it was established in section \ref{sec:pert_SW_prep}, the most general perturbative solution of (\ref{eq:invCon}) is
\eqn{i\neq j:\quad\frac{\partial^2}{\partial a_i\partial a_j}\mathcal{F}^{\textrm{Dell,pert}}=-\frac1{2\pi\imath}
\log\brc{1-\frac{m^2}{\sn\brc{\beta\brc{a_i-a_j}|\,\hat{\tau}}^2}},\quad
\frac{\partial^2}{\partial a_i^2}\mathcal{F}^{\textrm{Dell,pert}}=-
\sum_{\substack{j=1 \\ j\neq i}}^{N}
\frac{\partial^2}{\partial a_i\partial a_j}\mathcal{F}^{\textrm{Dell,pert}},}
where
\eq{\sn\brc{z|\,\hat{\tau}}=
\frac{\theta_{00}\brc{0|\,\hat{\tau}}}{\theta_{10}\brc{0|\,\hat{\tau}}}
\frac{\theta_{11}\brc{\hat{z}|\,\hat{\tau}}}{\theta_{01}\brc{\hat{z}|\,\hat{\tau}}},\quad
\hat{z}=\frac{z}{\pi\,\theta_{00}\brc{0|\,\hat{\tau}}^2}.}

We compute the instanton part of the $6d$ prepotential by solving the non-perturbative equations arising in the expansion of (\ref{eq:invCon}) in powers of $q$. The most general solutions of these equations are symmetric functions of $\bfi{\sn\brc{\beta\brc{a_i-a_j}|\,\hat\tau}}$ and, in the first non-trivial case of three particles, the instanton part of the $6d$ prepotential can be written in terms of the following variables:
\eq{
\label{eq:svar_Dell}
\begin{array}{c}
\hat{s}_{ij}\brc{\textbf{a},\beta,\hat{\tau}}=\frac14\left(\sn\brc{\beta\,a_{12}|\,\hat{\tau}}^{2i} \sn\brc{\beta\,a_{13}|\,\hat{\tau}}^{2j}+
\sn\brc{\beta\,a_{12}|\,\hat{\tau}}^{2i} \sn\brc{\beta\,a_{23}|\,\hat{\tau}}^{2j}+\right.\\
\left. +\sn\brc{\beta\,a_{13}|\,\hat{\tau}}^{2i}\sn\brc{\beta\,a_{23}|\,\hat{\tau}}^{2j}+
\brc{i\leftrightarrow j}
\right),\\
\hat{t}\brc{\textbf{a},\beta,\hat{\tau}}=\sn\brc{\beta\,a_{12}|\,\hat{\tau}}^2 \sn\brc{\beta\,a_{13}|\,\hat{\tau}}^2
\sn\brc{\beta\,a_{23}|\,\hat{\tau}}^2.
\end{array}}
The method of undetermined coefficients described in section \ref{sec:Inst3} works for the
double-elliptic system with slight modifications: the undetermined coefficients $c_{\dots}$ acquire the series expansion in powers of $m$ and the new parameter $\hat q$ (which is associated with the Kaluza-Klein compactification torus in the 5-th and 6-th dimensions),
\eq{\hat{q}=\frac{\theta_{10}\brc{0|\,\hat{\tau}}^4}{\theta_{00}\brc{0|\,\hat{\tau}}^4}.}
This allows us to write the instanton part of the prepotential $\mathcal{F}^{\textrm{Dell}}$ for $N=3$ as
\eqlb{eq:Dell,inst}{\boxed{
\mathcal{F}^{\textrm{Dell,inst}}=
\frac{m^2}{\pi\imath\,\beta^2}\sum_{n\in\mathbb{N}}\,\sum_{i=0}^{1}\,
\sum_{j=i}^{n}\,m^{6n-2i n-2j}\,\hat{c}_{n,in,j}\brc{m,\tau,\hat{\tau}}\,
\frac{\hat{s}_{in,j}}{\hat{t}^{n}},}}
where the functions $\hat{c}_{n,i,j}\brc{m,\tau,\hat{\tau}}$ are given by the series expansions
\eqlb{eq:c_Dell}{\hat{c}_{n,i,j}\brc{m,\tau,\hat{\tau}}=\sum_{l\geq\brc{n+1}/2}\, \sum_{k=0}^{{\max}_k}\,
\sum_{s=0}^{{\max}_s}\,m^{2k}\, \hat{q}^{s}\,\hat{c}_{n,i,j,k,s,l}\,q^l}
with rational coefficients $\hat{c}_{n,i,j,k,s,l}$ and the following notation:
\eq{{\max}_k=\min\bsq{3l+i+j-3n-1,\,4l-2n-2},\qq
{\max}_s=\min[k,\,2l-n-1].}
However, the latter formulas do not fully describe summation over the indices $k$ and $s$, since at higher orders in $q$ some coefficients $\hat{c}_{n,i,j,k,s,l}$ are systematically vanish.

We calculated the first $6$ instanton corrections in the form (\ref{eq:Dell,inst}) and the coefficients in the corresponding expansions (\ref{eq:c_Dell}) once again suggest the connection of the functions $\hat{c}_{n,i,j}\brc{m,\tau,\hat{\tau}}$ with the ring of quasimodular forms. As in the $5d$ case, the connection is more transparent with the proper choice of the mass parameter. In the $6d$ case, the natural mass parameter $\epsilon$ is related to the parameter $m$ in the following way:
\eq{m=\sn\brc{\epsilon|\,\hat\tau}.}
Rewriting the whole $6d$ prepotential as a series in $\epsilon$, we get
\eqlb{eq:prepDell_full}{\begin{array}{c}\ds
\mathcal{F}^{\textrm{Dell}}=\frac12\tau\sum_{i=1}^{3}a_i^2 -
\frac1{2\pi\imath}\sum_{i<j}\brc{a_{ij}}^2
\log\frac{\theta_{01}\brc{\hat\epsilon|\,\hat\tau}}
{\theta_{01}\brc{0|\,\hat\tau}}
+\frac{\epsilon^2}{4\pi\imath\,\beta^2}
\sum_{i<j}\log\,\theta_{11}\brc{\hat\beta\,a_{ij}|\,\hat\tau}^2+\\
\ds
+\frac{\epsilon^2}{\pi\imath\,\beta^2}\sum_{n\in\mathbb{N}}\,\sum_{i=0}^{1}\,
\sum_{j=i}^{n}\,\epsilon^{6n-2in-2j}\,C_{n,in,j}\brc{\epsilon,\tau,\hat\tau}\,
\frac{\hat s_{in,j}}{\hat t^{n}},
\end{array}}
where
\eq{\hat\epsilon=\frac{\epsilon}{\pi\,\theta_{00}\brc{0|\,\hat{\tau}}^2},\quad
\hat\beta=\frac{\beta}{\pi\,\theta_{00}\brc{0|\,\hat{\tau}}^2}}
and
\eqlb{eq:C_Dell}{C_{n,i,j}\brc{\epsilon,\tau,\hat\tau}
=\sum_{k=0}^{+\infty}\,\epsilon^{2k}\, C_{n,i,j,k}\brc{\tau,\hat\tau}
=\sum_{k=0}^{+\infty}\sum_{s=0}^{k}\,\epsilon^{2k}\,\hat{q}^s\, C_{n,i,j,k,s}\brc{\tau}.}

Using the computed coefficients in the expansions (\ref{eq:c_Dell}), we determine the exact expressions for a few first functions $C_{n,i,j,k}\brc{\tau,\hat\tau}$:
\eq{C_{1110}=-\frac{E_2}{12},\qq
C_{1111}=\frac{1+\hat{q}}{360}\brc{5\,E^2_2-E_4},}
\eqlb{eq:C1112}{C_{1112}=\frac{1+\hat{q}^2}{22\,680}\brc{-35\,E_2^3+21\,E_2 E_4-4\,E_6}
+\hat{q}\,\brc{\frac{-E_2^2+E_4}{288}-\frac{245\,E_2^3- 21\,E_2\,E_4 + 10\,E_6}{45\,360}},}
\eq{C_{1010}=\frac1{288}\brc{E_2^2-E_4},\qq
C_{1011}=-\frac{1+\hat{q}}{12\,960}\brc{25\,E_2^3-33\,E_2\,E_4+8\,E_6},}
\eq{C_{2220}=-\frac1{360}\brc{5\,E_2^2+E_4},\qq
C_{2221}=
\frac{1+\hat{q}}{45\,360}\brc{245\,E_2^3+42\,E_2\,E_4-17\,E_6}}
and so on. Clearly, there is no point in writing down all the terms in the series (\ref{eq:C_Dell}), since some of the functions $C_{n,i,j,k,s}\brc{\tau}$ coincide with the $5d$ functions $\tilde{c}_{n,i,j,k}\brc{\tau}$.

First of all, the degeneration of the $6d$ prepotential to the $5d$ one provides us with the following relations:
\eq{C_{n,i,j,k,0}\brc{\tau}=
\brc{-1}^k \tilde{c}^{\brc{\textrm{5d}}}_{n,i,j,k}\brc{\tau}}
and the results for $\tilde{c}_{n,i,j,k}\brc{\tau}$ can be seen in Appendix \ref{sec:5d_c}.
Another simple set of relations is given by direct computation of the functions $C_{n,i,j,k,k}\brc{\tau}$ up to weight $12$:
\eqlb{eq:C_rel}{C_{n,i,j,k,k}\brc{\tau}=
C_{n,i,j,k,0}\brc{\tau}=
\brc{-1}^k \tilde {c}^{\brc{\textrm{5d}}}_{n,i,j,k}\brc{\tau}.}
Thus, the only new functions arising in the $6d$ prepotential compared to the $5d$ one are $C_{n,i,j,k,s}\brc{\tau}$ with $0<s<k$. The first example of such a function is $C_{11121}\brc{\tau}$, and it is written down in (\ref{eq:C1112}) as
\eq{C_{11121}\brc{\tau}=\frac{-E_2^2+E_4}{288}-\frac{245\,E_2^3- 21\,E_2\,E_4 + 10\,E_6}{45\,360}.}
In general, the functions $C_{n,i,j,2,1}\brc{\tau}$ have the following form:
\eq{C_{n,i,j,2,1}\brc{\tau}=
\frac1{4\pi\imath}\frac{\partial}{\partial\tau} C_{n,i,j,0,0}\brc{\tau}+
\brc{\at{\textrm{quasimodular form of weight}}{6n-2i-2j+4}}.}
As we have seen in the $5d$ case, combinations of quasimodular forms with different weights appear, when some of the parameters in the prepotential are not appropriately chosen.
The form of the functions $C_{n,i,j,2,1}\brc{\tau}$ imply that the parameter $\tau$ in the $6d$ prepotential (\ref{eq:prepDell_full}) should be shifted by some function of $\epsilon$ and $\hat\tau$:
\eq{\tau=\tau' -\frac1{4\pi\imath}\hat{q}\,\epsilon^4+O\brc{\epsilon^6}.}
Computing the exact expressions for other functions $C_{n,i,j,k,s}\brc{\tau}$ with $k>2$, $0<s<k$, we determine the shift of the parameter $\tau$ as
\eq{\tau=\tau'+\frac3{\pi\imath}\brc{
\log\frac{\theta_{01}\brc{\hat\epsilon|\,\hat\tau}}
{\theta_{01}\brc{0|\,\hat\tau}}
-\frac12\,\vartheta\brc{\hat\tau}\,\epsilon^2},\qq
\vartheta\brc{\hat\tau}\equiv\frac{4\imath}{\pi}
\frac{\partial_{\hat\tau}\log\theta_{01}\brc{0|\,\hat\tau}}
{\theta_{00}\brc{0|\,\hat\tau}^4}.}

Finally, we obtain the proper set of parameters for the $6d$ prepotential, which is
$\epsilon$, $\beta$, $\tau'$ and $\hat\tau$. Then, the series expansion in powers of $\epsilon$ for the prepotential $\mathcal{F}^{\textrm{Dell}}$ acquires the form
\eqlb{eq:prepDell_fin}{\boxed{\begin{array}{c}\ds
\mathcal{F}^{\textrm{Dell}}=\frac12\sum_{i=1}^{3}a_i^2
\brc{\tau'+\frac3{\pi\imath}
\log\frac{\theta_{01}\brc{\hat\epsilon|\,\hat\tau}}
{\theta_{01}\brc{0|\,\hat\tau}}
-\frac{3}{2\pi\imath}\,\vartheta\brc{\hat\tau}\,\epsilon^2}
-\frac1{2\pi\imath}\sum_{i<j}\brc{a_{ij}}^2
\log\frac{\theta_{01}\brc{\hat\epsilon|\,\hat\tau}}
{\theta_{01}\brc{0|\,\hat\tau}}+\\
\ds
+\frac{\epsilon^2}{4\pi\imath\,\beta^2}
\sum_{i<j}\log\,\theta_{11}\brc{\hat\beta\,a_{ij}|\,\hat\tau}^2
+\frac{\epsilon^2}{\pi\imath\,\beta^2}\sum_{n\in\mathbb{N}}\,\sum_{i=0}^{1}\,
\sum_{j=i}^{n}\,\epsilon^{6n-2in-2j}\,\widehat{C}_{n,in,j}\brc{\epsilon,\tau',\hat\tau}\,
\frac{\hat s_{in,j}}{\hat t^{n}},
\end{array}}}
where
\eqlb{eq:hatC_Dell}{\widehat{C}_{n,i,j}\brc{\epsilon,\tau',\hat\tau}
=\sum_{k=0}^{+\infty}\sum_{s=0}^{k}\,\epsilon^{2k}\,\hat{q}^s\,
\widehat{C}_{n,i,j,k,s}\brc{\tau'}}
and the functions $\widehat{C}_{n,i,j,k,s}\brc{\tau'}$ are the quasimodular forms of weight $6n-2i-2j+2k$. The first $6$ instanton corrections allow us to obtain the exact expressions for $\widehat{C}_{n,i,j,k,s}\brc{\tau'}$ up to weight $12$. Since the change of the parameter $\tau$ does not affect the relations (\ref{eq:C_rel}),
\eq{\widehat{C}_{n,i,j,k,k}\brc{\tau'}=
\widehat{C}_{n,i,j,k,0}\brc{\tau'}=
\brc{-1}^k \tilde {c}^{\brc{\textrm{5d}}}_{n,i,j,k}\brc{\tau'},}
we only need to write down the results for the functions $\widehat{C}_{n,i,j,k,s}\brc{\tau'}$ with
$0<s<k$, which is done in Appendix \ref{sec:6d_C}.


\section{Conclusion}
We proposed new non-linear equations that allow one to effectively describe the instanton expansions for the Seiberg-Witten prepotentials associated with the $N=3$ elliptic Calogero-Moser system ($4d$ case), the $N=3$ elliptic Ruijsenaars system ($5d$ case) and the $N=3$ double-elliptic integrable system ($6d$ case). All the instanton expansions can be written in a universal manner as expansions in powers of flat moduli (including the adjoint matter hypermultiplet mass) with coefficients that are quasimodular forms of the elliptic parameter. Although the results are given for the first non-trivial case of $N=3$, the generalization to the case of $N>3$ is straightforward.
To obtain the instanton expansions for the general case with $N>3$, one should replace the
$N=3$ functions $s_{ij}\brc{\textbf{a}}$, $t\brc{\textbf{a}}$ (\ref{eq:svar_CM}), (\ref{eq:svar_RS}), (\ref{eq:svar_Dell}) by other sets of functions $s_{i_1 i_2\dots i_n}\brc{\textbf{a}}$, $t\brc{\textbf{a}}$ depending on all differences
$\brc{a_i-a_j}$, $i,j=1,\dots, N$.

An interesting problem is to describe the modular properties of the $6d$ spectral curve corresponding to the double-elliptic system.
The quasimodular properties of the coefficients in the expansion (\ref{eq:prepDell_fin}) of the $6d$ prepotential suggest that, similarly to the $4d$ and $5d$ cases (eqs.(\ref{eq:Mod_Eq_CM}) and (\ref{eq:Eq_Mod_RS}) respectively), there exists a $6d$ modular anomaly equation. We are going to address this problem elsewhere \cite{AMM}.

\section*{Acknowledgements}

Our work is partly supported by RFBR grants 15-02-04175 (G.A.), 16-01-00291 (A.Mir.), 16-02-01021 (A.Mor.), by grants mol-a-ved 15-31-20484 (G.A.), mol-a-ved 15-31-20832 (A.Mor),  mol-a 16-32-00920 (G.A.) and by joint grants 15-51-52031-NSC-a, 15-52-50041-YaF and 16-51-53034-GFEN.

\appendix

\section{Recurrence relations for the coefficients $e_n$}
\label{sec:recf}
In this Appendix, we present recurrence relations for the coefficients in the series expansion
\eq{f\brc{x}=x^2 + \sum_{n=2}^{\infty}e_{n-1}\,x^{2n}.}
For the first $10$ coefficients, these are
\eqlb{eq:e4B}{e_4=\frac{2}{3}e_1^4 - \frac{7}{3} e_1^2\, e_2 +2 e_1\, e_3+ \frac{2}{3} e_2^2,}
\eq{e_5= \frac{20}{33}e_1^5-\frac{49}{33} e_1^3\, e_2 + \frac{14}{11} e_1^2\, e_3 - \frac{37}{33} e_1\, e_2^2+\frac{19}{11} e_2\, e_3,}
\eq{e_6=\frac{12}{143}e_1^6+\frac{118}{143}e_1^4\, e_2 +\frac{56}{143}e_1^3\, e_3 -\frac{482}{143} e_1^2\,e_2^2 +\frac{252}{143}e_1\, e_2\, e_3 +\frac{5}{13}e_2^3+\frac{12}{13}e_3^2,}
\eq{e_7=\frac{1}{13}e_1^7-\frac{37}{143}e_1^5\, e_2+\frac{53}{39}e_1^4\, e_3+\frac{89}{143} e_1^3\, e_2^2 -\frac{1673}{429}e_1^2\, e_2\, e_3 -\frac{228}{143}e_1\, e_2^3 +\frac{37}{13} e_1\, e_3^2 +\frac{61}{33} e_2^2\, e_3,}
\eqn{e_8= \frac{62}{153}e_1^8-\frac{5354}{1989}e_1^6\, e_2+\frac{1382}{561}e_1^5\, e_3 +\frac{3440}{663}e_1^4\, e_2^2 -\frac{52402}{7293}e_1^3\,e_2\, e_3+\frac{644}{187}e_1^2\, e_3^2 -\frac{4652}{1989}e_1^2\, e_2^3-}
\eq{-\frac{5956}{7293}e_1\,e_2^2\, e_3+\frac{410}{1989} e_2^4+\frac{5653}{2431}e_2\, e_3^2,}
\eqn{e_9=\frac{3992}{10659}e_1^9-\frac{19667}{10659}e_1^7\,e_2+\frac{79828}{46189}e_1^6\, e_3 +\frac{2069}{3553}e_1^5\, e_2^2 -\frac{79191}{46189}e_1^4\,e_2\, e_3+\frac{56044}{10659}e_1^3\, e_2^3 +\frac{105840}{46189}e_1^3\, e_3^2-}
\eq{-\frac{527688}{46189}e_1^2\, e_2^2\, e_3+\frac{224028}{46189}e_1\,e_2\, e_3^2-\frac{16282}{10659}e_1\, e_2^4 +\frac{210}{247} e_3^3+\frac{4284}{2717}e_2^3\,e_3,}
\eqn{e_{10}=\frac{3160}{29887}e_1^{10}+\frac{392728}{1524237}e_1^8\,e_2 +\frac{71276}{89661}e_1^7\, e_3 -\frac{6909850}{1524237}e_1^6\, e_2^2+\frac{1149376}{1524237}e_1^5\,e_2\,e_3+\frac{4346533}{508079}e_1^4\,e_2^3+}
\eqlb{eq:e10B}{\frac{83680}{29887}e_1^4\,e_3^2-\frac{12758756}{1524237}e_1^3\,e_2^2\, e_3- \frac{1141770}{508079}e_1^2\,e_2\, e_3^2-\frac{642712}{1524237}e_1^2\,e_2^4+\frac{876}{247}e_1\, e_3^3-}
\eqn{-\frac{1915216}{508079}e_1\,e_2^3\,e_3+\frac{70}{663}e_2^5+\frac{91536}{26741}e_2^2\,e_3^2.}

\section{Instanton corrections from the spectral curve (\ref{eq:CMcurve})}
\label{sec:ECMcurve}
This Appendix is based on the paper \cite{HPh'98} by E. D’Hoker and D.H. Phong, where the first and the second instanton corrections were calculated.
We calculate here the first $4$ instanton corrections with the help of the spectral curve (\ref{eq:CMcurve}).

The spectral curve of the elliptic Calogero-Moser system can be rewritten in the form:
\eq{\sum_{n\in\mathbb{Z}}(-1)^{n}q^{\frac12n(n-1)}\rme^{nz}H\brc{k-mn}=0,}
where
\eq{H\brc{k}=\prod_{i=1}^{N}(k-k_i)}
and \eqs{k_i} are classical order parameters in Seiberg-Witten theory.

Introducing the new variable $y$:
\eqlb{eq:varzy}{\rme^{z}=y\frac{H(k)}{H(k-m)},}
we get the following equation for the curve:
\eqlb{eq:y}{y=1+\sum_{n=1}^{\infty}(-1)^{n}q^{\frac12n(n+1)}\brc{y^{-n}\eta_{n}\brc{k,1}-y^{n+1}\eta_{n}\brc{k-m,-1}}}
with
\eq{\eta_n\brc{k,\beta}=\frac{H\brc{k+\beta m n}H\brc{k-\beta m}^{n}}{H\brc{k}^{n+1}}.}

The classical and perturbative parts of the prepotential are well-known:
\eqn{\mathcal{F}^{\textrm{CM}}=\frac12\tau\sum_{i=1}^{N}a_i^2 + \frac12m \tau\sum_{i=1}^{N}a_i-} \eq{-\frac1{8\pi\imath}\sum_{i,j=1}^{N}\brc{\brc{a_i-a_j}^2\log\brc{a_i-a_j}^2-\brc{a_i-a_j+m}^2\log\brc{a_i-a_j+m}^2}+ \sum_{k\in\mathbb{N}}q^k \mathcal{F}^{(k)}.}

The simplest way to calculate the instanton corrections \eqs{\mathcal{F}^{(k)}} is to use the following equations:
\eqlb{eq:intai}{a_i=\frac1{2\pi\imath}\oint_{A_i}k \rmd z,}
\eqlb{eq:dFapp}{\frac{\partial \mathcal{F}^{\textrm{CM}}}{\partial\tau}=\frac1{4\pi\imath}\sum_{i=1}^{N}\oint_{A_i}k^2 \rmd z= \frac12\sum_{i=1}^{N}{k_i\brc{\textbf{a}}^2}+\textrm{const},}
where the integration contours are taken around the cuts (including the points $k_i$) on the corresponding sheets. The differentials can be rewritten as:
\eq{k \rmd z=k\rmd\log\brc{\frac{H\brc{k}}{H\brc{k-m}}}-\log\brc{y}\rmd k+ \rmd\brc{k\log\brc{y}},}
\eq{k^2 \rmd z=k^2\rmd\log\brc{\frac{H\brc{k}}{H\brc{k-m}}}-2k\log\brc{y}\rmd k+ \rmd\brc{k^2\log\brc{y}}.}
Since one can choose the integration contours located at a finite fixed distance from the points $k_i$, the functions $\eta_n$ in the integrand remain finite as $q\rightarrow 0$, and equation (\ref{eq:y}) can be used. The instanton corrections are calculated by the residue methods only \cite{HPh'98}, and the answer depends on the functions $S_{n,i}\brc{k}$ and $P_{n,i}\brc{k}$:
\eq{\eta_n\brc{k,1}=\frac1{\brc{k-k_i}^{n+1}}S_{n,i}\brc{k},\quad \eta_n\brc{k-m,-1}=\brc{k-k_i}^{n}P_{n,i}\brc{k},}
\eqlb{eq:defSP}{S_{n,i}\brc{k}=\frac{H\brc{k+ m n}H\brc{k-m}^{n}}{H_i\brc{k}^{n+1}},\quad
P_{n,i}\brc{k}=\frac{H_i\brc{k}^{n}H\brc{k-m\brc{n+1}}}{H\brc{k-m}^{n+1}},
}
where
\eq{H_i\brc{k}=\prod_{j\neq i}(k-k_j).}

Computing $\log\brc{y}$ up to the fourth order in $q$, one gets
\eqn{\log\brc{y}=q \left(\eta _{1,-1}-\eta _{1,1}\right)
+\frac{3}{2} q^2 \left(\eta _{1,-1}^2-\eta _{1,1}^2\right)+}
\eq{+\frac{1}{3} q^3 \left(10 \eta _{1,-1}^3 +3 \eta _{2,1}+3 \eta _{1,1}^2 \eta _{1,-1}-10 \eta _{1,1}^3-3 \eta _{2,-1} -3 \eta _{1,1} \eta _{1,-1}^2\right)+}
\eqn{+\frac{1}{4} q^4 \left(35 \eta _{1,-1}^4+ 16 \eta _{1,1} \eta _{2,1} +16 \eta _{1,1}^3 \eta _{1,-1}+ 4 \eta _{1,1} \eta _{2,-1}-\right.}
\eqn{\left. -35 \eta _{1,1}^4 -16 \eta _{1,1} \eta _{1,-1}^3-16 \eta _{2,-1} \eta _{1,-1}-4 \eta _{2,1} \eta _{1,-1}\right)+ O\brc{q^5},}
where \eqs{\eta _{n,1}=\eta_n\brc{k,1}} and \eqs{\eta _{n,-1}=\eta_n\brc{k-m,-1}}. In the first and the second orders, only the function $\eta _{1,1}$ exhibits poles at $k=k_i$. Thus, in the first and the second instanton orders, only the function $S_{1,i}\brc{k}$ appears. In the third and the fourth orders, both the functions $S_{n,i}\brc{k}$ and $P_{n,i}\brc{k}$ contribute, but the second one appears in the following combinations only:
\eq{S_{1,i}\left(k\right)^2\,P_{1,i}\left(k\right)=\frac{H\brc{k+m}^2 H\brc{k-2m}} {H_i\brc{k}^{3}},}
\eq{S_{1,i}\left(k\right)^3\,P_{1,i}\left(k\right)=\frac{H\brc{k+m}^3 H\brc{k-m} H\brc{k-2m}} {H_i\brc{k}^{5}},}
\eq{S_{2,i}\left(k\right)\,P_{1,i}\left(k\right)=\frac{H\brc{k+2m} H\brc{k-2m}}{H_i\brc{k}^{2}},}
which do not contain the "anomalous" poles at the points $k=k_i+m$. This means that the "anomalous" poles of the functions $P_{n,i}\brc{k}$ cancel out in the expressions for all relevant quantities.
In particular, none of the functions $k_i\brc{\textbf{a}}$ in (\ref{eq:dFapp}) contain "anomalous" poles, as well as the instanton corrections (at least up to the fourth order).

Now we use (\ref{eq:intai}) to calculate $a_i$'s:
\eqn{a_i=k_i+q\, S_{1,i}'\brc{k_i}+\frac{q^2}4\,\frac{\partial^3}{\partial k^3} \left.\brc{S_{1,i}\brc{k}^2}\right|_{k=k_i}+
\frac{q^3}{36}\,\frac{\partial^5}{\partial k^5}
\left.\brc{S_{1,i}\brc{k}^3}\right|_{k=k_i}-}
\eqlb{eq:ai}{-\frac{q^3}2\,\frac{\partial^2}{\partial k^2}
\left.\brc{S_{2,i}\brc{k}+ S_{1,i}\left(k\right)^2\,P_{1,i}\left(k\right)}\right|_{k=k_i}+
\frac{q^4}{576}\,\frac{\partial^7}{\partial k^7}
\left.\brc{S_{1,i}\brc{k}^4}\right|_{k=k_i}-}
\eqn{-\frac{q^4}6\, \frac{\partial^4}{\partial k^4}
\left.\brc{S_{1,i}\brc{k}\, S_{2,i}\brc{k}+
S_{1,i}\brc{k}^3\,P_{1,i}\brc{k}}\right|_{k=k_i}+
q^4\, \frac{\partial}{\partial k}
\left.\brc{S_{2,i}\brc{k}\,P_{1,i}\brc{k}}\right|_{k=k_i}
+O\brc{q^5}.}
Then, inverting the dependence in (\ref{eq:ai}), we obtain the first $4$ instanton corrections. Before writing them down, we would like to notice that in the three-particle case ($N=3$) all corrections agree with the expansion (\ref{eq:prepCM}). The explicit expressions for the first $3$ corrections are
\eq{\mathcal{F}^{(1)}=\frac1{2\pi\imath}\sum_{i=1}^{N}S_{1,i}\brc{a_i},}
\eq{\mathcal{F}^{(2)}=\frac1{4\pi\imath}\sum_{i=1}^{N}\brc{ S_{1,i}'\brc{a_i}^2+\frac32S_{1,i}\brc{a_i}S_{1,i}''\brc{a_i}- \sum_{j=1}^{N}S_{1,j}'\brc{a_j}\frac{\partial}{\partial a_j}S_{1,i}\brc{a_i}},}
\eq{\mathcal{F}^{(3)}=\frac{1}{72\pi\imath} \sum _{i=1}^{N}\left(S_{1,i}\left(a_i\right)^2\left(5 S_{1,i}''''\left(a_i\right)-12\,P_{1,i}'\left(a_i\right)\right)- 18\,S_{1,i}\brc{a_i} \sum_{j=1}^{N}S_{1,j}'\brc{a_j}\frac{\partial}{\partial a_j}S_{1,i}''\brc{a_i}\right)+}
\eqn{+\frac{1}{72\pi\imath} \sum _{i=1}^{N}\left(S_{1,i}\left(a_i\right) \left(-24 P_{1,i}\left(a_i\right) S_{1,i}'\left(a_i\right)+30 S_{1,i}''\left(a_i\right)^2+34 S_{1,i}'\left(a_i\right) S_{1,i}'''\left(a_i\right)\right)\right)-}
\eqn{-\frac{1}{12\pi\imath}\sum _{i,j=1}^{N}\left(4 S_{1,i}'\left(a_i\right) S_{1,j}'\left(a_j\right)\frac{\partial S_{1,i}'\left(a_i\right)}{\partial a_j}+3 S_{1,j}'\left(a_j\right) \left(S_{1,i}''\left(a_i\right)+S_{1,j}''\left(a_j\right)\right) \frac{\partial S_{1,i}\left(a_i\right)}{\partial a_j}\right)-}
\eqn{-\frac{1}{12\pi\imath}\sum _{i=1}^{N} \left(2 S_{2,i}'\left(a_i\right)-7 S_{1,i}'\left(a_i\right)^2 S_{1,i}''\left(a_i\right)+ \sum_{j=1}^{N}S_{1,j}\left(a_j\right) S_{1,j}'''\left(a_j\right) \frac{\partial S_{1,i}\left(a_i\right)}{\partial a_j}\right)+}
\eqn{+\frac{1}{12\pi\imath}\sum _{i,j,k=1}^{N}\left(S_{1,j}'\left(a_j\right) S_{1,k}'\left(a_k\right) \frac{\partial^2 S_{1,i}\left(a_i\right)}{\partial a_j\partial a_k}+2 S_{1,k}'\left(a_k\right)\frac{\partial S_{1,i}\left(a_i\right)}{\partial a_j} \frac{\partial S_{1,j}'\left(a_j\right)}{\partial a_k}\right),}
where we insert $a_i$'s instead of $k_i$'s in all functions $S_{n,i}\brc{a}$ and $P_{n,i}\brc{a}$. The expression for the fourth instanton correction is too long, so we just use equation (\ref{eq:dFapp}) and compare the result with (\ref{eq:prepCM}).

\section{$4d$ functions $c_{n,i,j}\brc{\tau}$}
Functions listed in table (\ref{eq:cTab}) are:
\label{sec:4d_c}
\eq{c_{111}=\frac1{12}-\frac{E_2}{12},}
\eq{c_{101}=\frac1{288}\brc{E_2^2-E_4},\qq
c_{222}=\frac1{60}-\frac1{360}\brc{5 E_2^2+E_4},}
\eq{c_{100} =\frac1{25\,920}\brc{-5 E_2^3+3 E_2 E_4+2 E_6},\quad
c_{221} =\frac1{2\,160} \brc{5 E_2^3-3 E_2 E_4-2 E_6},}
\eq{c_{333}=\frac1{168}+\frac1{45\,360}\brc{-175 E_2^3-84 E_2 E_4-11 E_6},}
\eq{c_{202}=\frac1{967\,680}\brc{175 E_2^4+14 E_2^2 E_4-85 E_4^2-104 E_2 E_6},}
\eq{c_{332}=\frac1{36\,288}\brc{35 E_2^4-7 E_2^2 E_4-10 E_4^2-18 E_2 E_6},}
\eq{c_{444}=\frac1{360}-\frac1{181\,440}\brc{245 E_2^4+196 E_2^2 E_4+19 E_4^2+44 E_2 E_6},}
\eq{c_{201}=\frac1{1\,451\,520}\brc{175 E_2^5-14 E_2^3 E_4-81 E_2 E_4^2-136 E_2^2 E_6+56 E_4 E_6},}
\eq{c_{331}=\frac1{2\,177\,280}\brc{175 E_2^5+203 E_2^3 E_4-174 E_2 E_4^2-43 E_2^2 E_6-161 E_4 E_6},}
\eq{c_{443}=\frac1{77\,760}\brc{35 E_2^5+7 E_2^3 E_4-18 E_2 E_4^2-17 E_2^2 E_6-7 E_4 E_6},}
\eq{c_{555}=\frac1{660}-
\frac1{9\,979\,200}\brc{5390 E_2^5+6160 E_2^3 E_4+1496 E_2 E_4^2+1815 E_2^2 E_6+259 E_4 E_6},}
\eq{c_{200}=-\frac{259 E_2^6}{1990656}-\frac{109 E_2^4 E_4}{3317760}+\frac{31609 E_2^2 E_4^2}{348364800}-\frac{6095 E_4^3}{459841536}+\frac{2729 E_2^3 E_6}{26127360}-\frac{1741 E_2 E_4 E_6}{479001600}-\frac{787 E_6^2}{51321600},}
\eq{c_{303}=\frac{455 E_2^6}{5971968}+\frac{85 E_2^4 E_4}{1990656}-\frac{11593 E_2^2 E_4^2}{209018880}-\frac{485 E_4^3}{1379524608}-\frac{185 E_2^3 E_6}{3919104}-
\frac{353 E_2 E_4 E_6}{17962560}+\frac{233 E_6^2}{61585920},}
\eq{c_{442}=\frac{7 E_2^6}{746496}+\frac{7 E_2^4 E_4}{124416}-\frac{11 E_2^2 E_4^2}{1741824}-
\frac{613 E_4^3}{28740096}+\frac{23 E_2^3 E_6}{870912}-\frac{851 E_2 E_4 E_6}{15966720}-\frac{19 E_6^2}{1710720},}
\eq{c_{554}=\frac{7 E_2^6}{31104}+\frac{7 E_2^4 E_4}{51840}-\frac{241 E_2^2 E_4^2}{1814400}-
\frac{61 E_4^3}{2395008}-\frac{23 E_2^3 E_6}{272160}-\frac{271 E_2 E_4 E_6}{2494800}-
\frac{37 E_6^2}{4276800},}
\eq{c_{666}=\frac1{1092}-\frac{11 E_2^6}{46656}-\frac{11 E_2^4 E_4}{31104}-
\frac{1199 E_2^2 E_4^2}{8164800}-\frac{2281 E_4^3}{280215936}-\frac{121 E_2^3 E_6}{979776}-
\frac{4127 E_2 E_4 E_6}{89812800}-\frac{3313 E_6^2}{1751349600},}
\eq{c_{302}=-\frac{4141 E_2^7}{23887872}-\frac{275 E_2^5 E_4}{2654208}+\frac{120487 E_2^3 E_4^2}{836075520}-\frac{701089 E_2 E_4^3}{119558799360}+\frac{10849 E_2^4 E_6}{83607552}+}
\eqn{+\frac{16033 E_2^2 E_4 E_6}{255467520}-\frac{159023 E_4^2 E_6}{5434490880}-
\frac{12227 E_2 E_6^2}{498161664},}
\eq{c_{441}=\frac{83 E_2^7}{995328}+\frac{649 E_2^5 E_4}{8957952}-\frac{18841 E_2^3 E_4^2}{313528320}-\frac{521861 E_2 E_4^3}{44834549760}-\frac{359 E_2^4 E_6}{7838208}-
\frac{4051 E_2^2 E_4 E_6}{86220288}+}
\eqn{+\frac{1201 E_4^2 E_6}{254741760}+\frac{823 E_2 E_6^2}{207567360},}
\eq{c_{553}=-\frac{E_2^7}{124416}+\frac{E_2^5 E_4}{41472}+\frac{67 E_2^3 E_4^2}{3110400}-
\frac{1829 E_2 E_4^3}{88957440}+\frac{7 E_2^4 E_6}{248832}-\frac{203 E_2^2 E_4 E_6}{11404800}-}
\eqn{-\frac{1759 E_4^2 E_6}{113218560}-\frac{18449 E_2 E_6^2}{1556755200},}
\eq{c_{665}=\frac{11 E_2^7}{93312}+\frac{11 E_2^5 E_4}{93312}-\frac{187 E_2^3 E_4^2}{3265920}-
\frac{1669 E_2 E_4^3}{42456960}-\frac{11 E_2^4 E_6}{435456}-\frac{283 E_2^2 E_4 E_6}{3265920}-}
\eqn{-\frac{1177 E_4^2 E_6}{84913920}-\frac{19 E_2 E_6^2}{1415232},}
\eq{c_{777}=\frac1{1680}-\frac{143 E_2^7}{1306368}-\frac{143 E_2^5 E_4}{699840}-
\frac{377 E_2^3 E_4^2}{3061800}-\frac{62459 E_2 E_4^3}{3502699200}-\frac{1573 E_2^4 E_6}{19595520}-
\frac{56797 E_2^2 E_4 E_6}{1077753600}-}
\eqn{\frac{7907 E_4^2 E_6}{2228990400}-\frac{43151 E_2 E_6^2}{10897286400}.}

\section{$5d$ functions $\tilde c_{n,i,j,k}\brc{\tau}$, $k>0$}
\label{sec:5d_c}
The functions with weights not greater than $14$ are
\eq{
\begin{array}{|c|ccccccc|}
\hline
\textrm{Weight} & \textrm{Functions} &&&&&&\\ \hline
4  & \tc_{1111} &            &            &            &            & &\\ \hline
6  & \tc_{1112} & \tc_{1011} & \tc_{2221} &            &            & &\\ \hline
8  & \tc_{1113} & \tc_{1012} & \tc_{2222} & \tc_{1001} & \tc_{2211} & \tc_{3331} &\\ \hline
10 & \tc_{1114} & \tc_{1013} & \tc_{2223} & \tc_{1002} & \tc_{2212} & \tc_{3332} &\\ \hline
12 & \tc_{1115} & \tc_{1014} & \tc_{2224} & \tc_{1003} & \tc_{2213} & \tc_{3333} &\\ \hline
14 & \tc_{1116} & \tc_{1015} & \tc_{2225} & \tc_{1004} & \tc_{2214} & \tc_{3334} &\\ \hline
&&&&&&&\\ \hline
10 & \tc_{2021} & \tc_{3321} & \tc_{4441} &            &            &            & \\ \hline
12 & \tc_{2022} & \tc_{3322} & \tc_{4442} & \tc_{2011} & \tc_{3311} & \tc_{4431} & \tc_{5551}\\ \hline
14 & \tc_{2023} & \tc_{3323} & \tc_{4443} & \tc_{2012} & \tc_{3312} & \tc_{4432} & \tc_{5552}\\ \hline
&&&&&&&\\ \hline
14 & \tc_{2001} & \tc_{3031} & \tc_{4421} & \tc_{5541} & \tc_{6661} &            & \\ \hline
\end{array}}

\eq{\tilde {c}_{1111}=\frac1{360}\brc{-5\,E^2_2+E_4},}
\eq{\tc_{1112}=\frac1{22\,680}\brc{-35 E_2^3+21 E_2 E_4-4 E_6}}
\eq{\tilde {c}_{1011}=\frac1{12\,960}\brc{25\,E_2^3-33\,E_2\,E_4+8\,E_6},}
\eq{\tilde {c}_{2221}=-
\frac1{45\,360}\brc{245\,E_2^3+42\,E_2\,E_4-17\,E_6},}
\eq{\tilde {c}_{1113}=\frac1{1360800} \brc{-175 E_2^4+210 E_2^2 E_4-80 E_2 E_6-3 E_4^2},}
\eq{\tilde {c}_{1012}=\frac1{1088640}\brc{595 E_2^4-1050 E_2^2 E_4+464 E_2 E_6-9 E_4^2},}
\eq{\tilde {c}_{2222}=\frac1{2721600}\brc{-3325 E_2^4-210 E_2^2 E_4+520 E_2 E_6-9 E_4^2},}
\eq{\tilde {c}_{2211}=\frac1{725760}\brc{1295 E_2^4-966 E_2^2 E_4-464 E_2 E_6+135 E_4^2},}
\eq{\tilde {c}_{1001}=-\frac1{20736}\brc{E_2^2-E_4}^2,\quad
\tilde {c}_{3331}=\frac1{272160}\brc{-665 E_2^4-357 E_2^2 E_4+2 E_2 E_6+12 E_4^2},}
\eq{\tc_{1114}=\frac1{44906400}\brc{-385 E_2^5+770 E_2^3 E_4-440 E_2^2 E_6-33 E_2 E_4^2+40 E_4 E_6},}
\eq{\tc_{1013}=\frac1{3265920}\brc{343 E_2^5-798 E_2^3 E_4+488 E_2^2 E_6-9 E_2 E_4^2-24 E_4 E_6},}
\eq{\tc_{2223}=\frac1{17962560}\brc{-3619 E_2^5+462 E_2^3 E_4+880 E_2^2 E_6-99 E_2 E_4^2-72 E_4 E_6},}
\eq{\tc_{1002}=\frac1{6531840}\brc{-483 E_2^5+182 E_2^3 E_4+264 E_2^2 E_6+93 E_2 E_4^2-56 E_4 E_6},}
\eq{\tc_{2212}=\frac{E_2}{2177280}\left(1547 E_2^4-1470 E_2^2 E_4-464 E_2 E_6+387 E_4^2\right),}
\eq{\tc_{3332}=\frac1{17962560}\brc{-15323 E_2^5-8316 E_2^3 E_4+1496 E_2^2 E_6+891 E_2 E_4^2+84 E_4 E_6},}
\eq{\tc_{2021}=\frac1{8709120}\brc{2275 E_2^5+266 E_2^3 E_4-1576 E_2^2 E_6-1245 E_2 E_4^2+280 E_4 E_6},}
\eq{\tc_{3321}=\frac1{3265920}\brc{3185 E_2^5-770 E_2^3 E_4-1730 E_2^2 E_6-867 E_2 E_4^2+182 E_4 E_6},}
\eq{\tc_{4441}=\frac1{17962560}\brc{-21560 E_2^5-19558 E_2^3 E_4-2827 E_2^2 E_6-1716 E_2 E_4^2+301 E_4 E_6},}
\eq{\tc_{1115}=-\frac{E_2^6}{2099520}+\frac{E_2^4 E_4}{699840}-\frac{E_2^3 E_6}{918540}-\frac{E_2^2 E_4^2}{8164800}+\frac{E_2 E_4 E_6}{3367980}-\frac{113 E_4^3}{3502699200}-\frac{34 E_6^2}{1149323175},}
\eq{\tc_{1014}=\frac{43 E_2^6}{2799360}-\frac{43 E_2^4 E_4}{933120}+\frac{43 E_2^3 E_6}{1224720}+\frac{E_2^2 E_4^2}{32659200}-\frac{17 E_2 E_4 E_6}{3207600}+\frac{113 E_4^3}{215550720}+\frac{E_6^2}{2806650},}
\eq{\tc_{2224}=-\frac{37 E_2^6}{1399680}+\frac{E_2^4 E_4}{93312}+\frac{E_2^3 E_6}{122472}-\frac{47 E_2^2 E_4^2}{16329600}-\frac{23 E_2 E_4 E_6}{11226600}+\frac{1921 E_4^3}{7005398400}+\frac{29 E_6^2}{153243090},}
\eq{\tc_{1003}=\frac{53 E_2^6}{1679616}+\frac{139 E_2^4 E_4}{2799360}-\frac{4 E_2^3 E_6}{229635}-\frac{4733 E_2^2 E_4^2}{97977600}-\frac{11 E_2 E_4 E_6}{255150}+\frac{23 E_4^3}{1679616}+\frac{16 E_6^2}{1148175},}
\eq{\tc_{2213}=\frac{181 E_2^6}{933120}-\frac{73 E_2^4 E_4}{311040}-\frac{17 E_2^3 E_6}{408240}+\frac{967 E_2^2 E_4^2}{10886400}-\frac{23 E_2 E_4 E_6}{7484400}-\frac{41 E_4^3}{14370048}-\frac{E_6^2}{1871100},}
\eq{\tc_{3333}=-\frac{11 E_2^6}{51840}-\frac{49 E_2^4 E_4}{466560}+\frac{37 E_2^3 E_6}{816480}+\frac{59 E_2^2 E_4^2}{2332800}+\frac{13 E_2 E_4 E_6}{44906400}-\frac{1577 E_4^3}{1401079680}-\frac{97 E_6^2}{153243090},}
\eq{\tc_{2022}=\frac{169 E_2^6}{1866240}-\frac{11 E_2^4 E_4}{622080}-\frac{197 E_2^3 E_6}{2449440}-\frac{2519 E_2^2 E_4^2}{65318400}+\frac{1201 E_2 E_4 E_6}{17962560}-\frac{785 E_4^3}{86220288}-\frac{7 E_6^2}{601425},}
\eq{\tc_{3322}=\frac{6853 E_2^6}{13436928}-\frac{733 E_2^4 E_4}{4478976}-\frac{1727 E_2^3 E_6}{5878656}-\frac{6779 E_2^2 E_4^2}{52254720}+\frac{1141 E_2 E_4 E_6}{15396480}+\frac{235 E_4^3}{49268736}-\frac{E_6^2}{577368},}
\eq{\tc_{4442}=-\frac{1201 E_2^6}{2099520}-\frac{781 E_2^4 E_4}{1399680}-\frac{457 E_2^3 E_6}{14696640}-\frac{247 E_2^2 E_4^2}{8164800}+\frac{7841 E_2 E_4 E_6}{269438400}+\frac{3121 E_4^3}{1401079680}+\frac{23 E_6^2}{262702440},}
\eq{\tc_{2011}=-\frac{209 E_2^6}{995328}-\frac{121 E_2^4 E_4}{995328}+\frac{391 E_2^3 E_6}{2612736}+\frac{32899 E_2^2 E_4^2}{174182400}+\frac{6203 E_2 E_4 E_6}{79833600}-\frac{9461 E_4^3}{229920768}-\frac{1117 E_6^2}{25660800},}
\eq{\tc_{3311}=\frac{1777 E_2^6}{8957952}+\frac{2707 E_2^4 E_4}{14929920}-\frac{2027 E_2^3 E_6}{19595520}-\frac{20141 E_2^2 E_4^2}{104509440}-\frac{8293 E_2 E_4 E_6}{71850240}+\frac{9995 E_4^3}{689762304}+\frac{67 E_6^2}{3849120},}
\eq{\tc_{4431}=\frac{637 E_2^6}{1119744}+\frac{119 E_2^4 E_4}{933120}-\frac{31 E_2^3 E_6}{103680}-\frac{583 E_2^2 E_4^2}{1866240}-\frac{23 E_2 E_4 E_6}{228096}+\frac{47 E_4^3}{6158592}+\frac{7 E_6^2}{855360},}
\eq{\tc_{5551}=-\frac{29 E_2^6}{46656}-\frac{31 E_2^4 E_4}{38880}-\frac{467 E_2^3 E_6}{2449440}-\frac{67 E_2^2 E_4^2}{326592}-\frac{5 E_2 E_4 E_6}{256608}+\frac{7 E_4^3}{10007712}+\frac{139 E_6^2}{70053984},}
\eq{\tc_{1116}=-\frac{E_2^7}{44089920}+\frac{E_2^5 E_4}{10497600}-\frac{E_2^4 E_6}{11022480}-\frac{E_2^3 E_4^2}{73483200}+\frac{E_2^2 E_4 E_6}{20207880}-\frac{113 E_2 E_4^3}{10508097600}-}
\eqn{-\frac{34 E_2 E_6^2}{3447969525}+\frac{229 E_4^2 E_6}{91945854000},}
\eq{\tc_{1015}=\frac{107 E_2^7}{58786560}-\frac{289 E_2^5 E_4}{41990400}+\frac{13 E_2^4 E_6}{2099520}+\frac{13 E_2^3 E_4^2}{58786560}-\frac{773 E_2^2 E_4 E_6}{404157600}+\frac{10451 E_2 E_4^3}{30023136000}+}
\eqn{+\frac{1151 E_2 E_6^2}{4597292700}-\frac{13421 E_4^2 E_6}{367783416000},}
\eq{\tc_{2225}=-\frac{17 E_2^7}{5878656}+\frac{47 E_2^5 E_4}{20995200}+\frac{E_2^4 E_6}{1049760}-\frac{127 E_2^3 E_4^2}{146966400}-\frac{101 E_2^2 E_4 E_6}{202078800}+\frac{18509 E_2 E_4^3}{105080976000}+}
\eqn{+\frac{487 E_2 E_6^2}{4597292700}-\frac{3527 E_4^2 E_6}{183891708000},}
\eq{\tc_{1004}=-\frac{1931 E_2^7}{50388480}-\frac{10823 E_2^5 E_4}{251942400}+\frac{383 E_2^4 E_6}{35271936}+\frac{36781 E_2^3 E_4^2}{1763596800}+\frac{30433 E_2^2 E_4 E_6}{692841600}+\frac{35302279 E_2 E_4^3}{1260971712000}+}
\eqn{+\frac{37507 E_2 E_6^2}{3940536600}-\frac{1826857 E_4^2 E_6}{57316896000},}
\eq{\tc_{2214}=\frac{797 E_2^7}{19595520}-\frac{217 E_2^5 E_4}{3499200}-\frac{37 E_2^4 E_6}{9797760}+\frac{601 E_2^3 E_4^2}{19595520}-\frac{793 E_2^2 E_4 E_6}{269438400}-\frac{90683 E_2 E_4^3}{35026992000}-}
\eqn{-\frac{1151 E_2 E_6^2}{3064861800}+\frac{7921 E_4^2 E_6}{22289904000},}
\eq{\tc_{3334}=-\frac{407 E_2^7}{9797760}-\frac{341 E_2^5 E_4}{20995200}+\frac{209 E_2^4 E_6}{14696640}+\frac{1177 E_2^3 E_4^2}{146966400}-\frac{289 E_2^2 E_4 E_6}{202078800}-\frac{98207 E_2 E_4^3}{105080976000}-}
\eqn{-\frac{263 E_2 E_6^2}{612972360}+\frac{4457 E_4^2 E_6}{33434856000},}
\eq{\tc_{2023}=\frac{32923 E_2^7}{470292480}+\frac{497 E_2^5 E_4}{13436928}-\frac{409 E_2^4 E_6}{8709120}-\frac{3041 E_2^3 E_4^2}{67184640}-\frac{61219 E_2^2 E_4 E_6}{6466521600}-\frac{6709037 E_2 E_4^3}{336259123200}-}
\eqn{-\frac{649237 E_2 E_6^2}{73556683200}+\frac{2511947 E_4^2 E_6}{106991539200},}
\eq{\tc_{3323}=\frac{51637 E_2^7}{282175488}-\frac{5383 E_2^5 E_4}{67184640}-\frac{1961 E_2^4 E_6}{17635968}-\frac{19511 E_2^3 E_4^2}{470292480}+\frac{457 E_2^2 E_4 E_6}{9237888}+\frac{1588183 E_2 E_4^3}{336259123200}-}
\eqn{-\frac{28579 E_2 E_6^2}{11033502480}-\frac{12013 E_4^2 E_6}{6686971200},}
\eq{\tc_{4443}=-\frac{8437 E_2^7}{44089920}-\frac{403 E_2^5 E_4}{2099520}+\frac{787 E_2^4 E_6}{88179840}+\frac{113 E_2^3 E_4^2}{73483200}+\frac{4033 E_2^2 E_4 E_6}{202078800}+\frac{8297 E_2 E_4^3}{5254048800}-}
\eqn{-\frac{6187 E_2 E_6^2}{8826801984}-\frac{9691 E_4^2 E_6}{13373942400},}
\eq{\tc_{2012}=\frac{7847 E_2^7}{14929920}+\frac{20009 E_2^5 E_4}{44789760}-\frac{24653 E_2^4 E_6}{78382080}-\frac{692411 E_2^3 E_4^2}{1567641600}-\frac{406517 E_2^2 E_4 E_6}{1077753600}-\frac{13699537 E_2 E_4^3}{224172748800}+}
\eqn{+\frac{50537 E_2 E_6^2}{1167566400}+\frac{911513 E_4^2 E_6}{5094835200},}
\eq{\tc_{3312}=\frac{1279 E_2^7}{14929920}+\frac{10403 E_2^5 E_4}{134369280}-\frac{3863 E_2^4 E_6}{58786560}-\frac{496289 E_2^3 E_4^2}{4702924800}-\frac{14009 E_2^2 E_4 E_6}{808315200}+\frac{20231069 E_2 E_4^3}{672518246400}+}
\eqn{+\frac{12319 E_2 E_6^2}{700539840}-\frac{84769 E_4^2 E_6}{3821126400},}
\eq{\tc_{4432}=\frac{923 E_2^7}{2488320}+\frac{5467 E_2^5 E_4}{67184640}-\frac{7157 E_2^4 E_6}{33592320}-\frac{74317 E_2^3 E_4^2}{335923200}-\frac{46087 E_2^2 E_4 E_6}{923788800}+\frac{163931 E_2 E_4^3}{9607403520}+}
\eqn{+\frac{701 E_2 E_6^2}{58378320}+\frac{8557 E_4^2 E_6}{3056901120},}
\eq{\tc_{5552}=-\frac{11 E_2^7}{29160}-\frac{5507 E_2^5 E_4}{10497600}-\frac{347 E_2^4 E_6}{3674160}-\frac{9797 E_2^3 E_4^2}{73483200}+\frac{3371 E_2^2 E_4 E_6}{404157600}+\frac{134819 E_2 E_4^3}{26270244000}+}
\eqn{+\frac{1231 E_2 E_6^2}{350269920}+\frac{1783 E_4^2 E_6}{1194102000},}
\eq{\tc_{2001}=\frac{841 E_2^7}{2985984}+\frac{623 E_2^5 E_4}{2488320}-\frac{17791 E_2^4 E_6}{104509440}-\frac{26533 E_2^3 E_4^2}{104509440}-\frac{41281 E_2^2 E_4 E_6}{191600640}-\frac{11083 E_2 E_4^3}{747242496}+}
\eqn{+\frac{21773 E_2 E_6^2}{622702080}+\frac{23753 E_4^2 E_6}{271724544},}
\eq{\tc_{3031}=-\frac{2933 E_2^7}{35831808}-\frac{5501 E_2^5 E_4}{107495424}+\frac{3349 E_2^4 E_6}{53747712}+\frac{40403 E_2^3 E_4^2}{537477120}+\frac{535907 E_2^2 E_4 E_6}{10346434560}-\frac{38239 E_2 E_4^3}{8277147648}-}
\eqn{-\frac{1229 E_2 E_6^2}{71850240}-\frac{12989 E_4^2 E_6}{376233984},}
\eq{\tc_{4421}=\frac{223 E_2^7}{2985984}+\frac{1303 E_2^5 E_4}{8957952}+\frac{67 E_2^4 E_6}{31352832}-\frac{3851 E_2^3 E_4^2}{62705664}-\frac{2005 E_2^2 E_4 E_6}{15676416}-\frac{161137 E_2 E_4^3}{4075868160}-}
\eqn{-\frac{155 E_2 E_6^2}{16982784}+\frac{31991 E_4^2 E_6}{2037934080},}
\eq{\tc_{5541}=\frac{E_2^7}{2916}+\frac{107 E_2^5 E_4}{466560}-\frac{1831 E_2^4 E_6}{13063680}-\frac{109 E_2^3 E_4^2}{510300}-\frac{62893 E_2^2 E_4 E_6}{359251200}-\frac{1931 E_2 E_4^3}{46702656}-}
\eqn{-\frac{16799 E_2 E_6^2}{2335132800}+\frac{173 E_4^2 E_6}{33965568},}
\eq{\tc_{6661}=-\frac{187 E_2^7}{559872}-\frac{583 E_2^5 E_4}{1049760}-\frac{10021 E_2^4 E_6}{58786560}-\frac{18307 E_2^3 E_4^2}{73483200}-\frac{210799 E_2^2 E_4 E_6}{3233260800}-\frac{40457 E_2 E_4^3}{2627024400}-}
\eqn{-\frac{193 E_2 E_6^2}{667180800}+\frac{263 E_4^2 E_6}{272937600}.}

\section{$6d$ functions $\widehat{C}_{n,i,j,k,s}\brc{\tau'}$, $0<s<k$}
\label{sec:6d_C}
Here we present the functions $\widehat{C}_{n,i,j,k,s}\brc{\tau'}$, $0<s<k$ up to weight $12$:
\eq{
\begin{array}{|c|ccc|}
\hline
\textrm{Weight} & \textrm{Functions} &&\\ \hline
6  & \wc_{11121} && \\ \hline
8  & \wc_{11131}=\wc_{11132} &&  \\ \hline
10 & \wc_{11141}=\wc_{11143},\quad \wc_{11142} && \\ \hline
12 & \wc_{11151}=\wc_{11154},\quad \wc_{11152}=\wc_{11153} && \\ \hline
&&&\\ \hline
8  & \wc_{10121} & \wc_{22221}&\\ \hline
10 & \wc_{10131}=\wc_{10132} & \wc_{22231}=\wc_{22232}&\\ \hline
12 & \wc_{10141}=\wc_{10143},\quad \wc_{10142} & \wc_{22241}=\wc_{22243},\quad \wc_{22242}&\\ \hline
&&&\\ \hline
10  & \wc_{10021} &  \wc_{22121} & \wc_{33321} \\ \hline
12 & \wc_{10031}=\wc_{10032} & \wc_{22131}=\wc_{22132} & \wc_{33331}=\wc_{33332}\\ \hline
&&&\\ \hline
12  & \wc_{20221} &  \wc_{33221} & \wc_{44421} \\ \hline
\end{array}}

The exact expressions are
\eq{\wc_{11121}=\frac1{45\,360}\brc{-245 E_2^3+21 E_2 E_4-10 E_6},}
\eq{\wc_{11131}=\wc_{11132}=\frac1{362\,880}\brc{385 E_2^4+42 E_2^2 E_4-3 E_4^2-40 E_2 E_6},}
\eq{\wc_{10121}=\frac1{4\,354\,560}\brc{8225 E_2^4-11046 E_2^2 E_4-99 E_4^2+2920 E_2 E_6},}
\eq{\wc_{22221}=\frac1{3\,110\,400}\brc{-11125 E_2^4-3450 E_2^2 E_4-9 E_4^2+760 E_2 E_6},}
\eq{\wc_{11141}=\wc_{11143}=
\frac1{179\,625\,600}\brc{-25795 E_2^5-13090 E_2^3 E_4-627 E_2 E_4^2+16060 E_2^2 E_6-644 E_4 E_6},}
\eq{\wc_{11142}=\frac1{479\,001\,600}\brc{-171325 E_2^5-133210 E_2^3 E_4-5313 E_2 E_4^2+95920 E_2^2 E_6-10328 E_4 E_6},}
\eq{\wc_{10131}=\wc_{10132}=
\frac1{4354560}\brc{-3115 E_2^5+5502 E_2^3 E_4-507 E_2 E_4^2-2084 E_2^2 E_6+204 E_4 E_6},}
\eq{\wc_{22231}=\wc_{22232}=\frac1{239500800}
\brc{295295 E_2^5+123970 E_2^3 E_4-10593 E_2 E_4^2-36080 E_2^2 E_6+1808 E_4 E_6},}
\eq{\wc_{10021}=\frac1{13063680}\brc{-1449 E_2^5+1064 E_2^3 E_4-159 E_2 E_4^2+474 E_2^2 E_6+70 E_4 E_6},}
\eq{\wc_{22121}=\frac1{2177280}\brc{4718 E_2^5-3283 E_2^3 E_4+297 E_2 E_4^2-1823 E_2^2 E_6+91 E_4 E_6},}
\eq{\wc_{33321}=\frac1{71850240}
\brc{-163933 E_2^5-114576 E_2^3 E_4-495 E_2 E_4^2-44 E_2^2 E_6+840 E_4 E_6},}
\eq{\wc_{11151}=\wc_{11154}=
\frac{E_2^6}{65610}+\frac{13 E_2^4 E_4}{699840}-\frac{29 E_2^3 E_6}{1049760}+\frac{E_2^2 E_4^2}{408240}+\frac{73 E_2 E_4 E_6}{19245600}-\frac{1061 E_4^3}{3502699200}-\frac{1601 E_6^2}{9194585400},}
\eq{\wc_{11152}=\wc_{11153}=
\frac{4427 E_2^6}{67184640}+\frac{1783 E_2^4 E_4}{22394880}-\frac{1859 E_2^3 E_6}{23514624}+\frac{8947 E_2^2 E_4^2}{261273600}+\frac{1873 E_2 E_4 E_6}{61585920}-}
\eqn{-\frac{98591 E_4^3}{12454041600}-\frac{1012603 E_6^2}{147113366400},}
\eq{\wc_{10141}=\wc_{10143}=\frac{1031 E_2^6}{3732480}-\frac{1297 E_2^4 E_4}{3732480}+\frac{187 E_2^3 E_6}{1399680}-\frac{E_2^2 E_4^2}{138240}-\frac{6043 E_2 E_4 E_6}{59875200}+\frac{431 E_4^3}{19160064}+\frac{449 E_6^2}{19245600},}
\eq{\wc_{10142}=
\frac{9229 E_2^6}{11197440}-\frac{359 E_2^4 E_4}{622080}-\frac{751 E_2^3 E_6}{78382080}-\frac{2923 E_2^2 E_4^2}{16329600}-\frac{39119 E_2 E_4 E_6}{205286400}+\frac{53369 E_4^3}{862202880}+\frac{1181 E_6^2}{16839900},}
\eq{\wc_{22241}=\wc_{22243}=
-\frac{103 E_2^6}{349920}-\frac{23 E_2^4 E_4}{155520}+\frac{41 E_2^3 E_6}{612360}+\frac{29 E_2^2 E_4^2}{816480}-\frac{131 E_2 E_4 E_6}{14968800}+}
\eqn{+\frac{223 E_4^3}{7005398400}+\frac{2507 E_6^2}{6129723600},}
\eq{\wc_{22242}=
-\frac{3751 E_2^6}{5971968}-\frac{12883 E_2^4 E_4}{29859840}+\frac{857 E_2^3 E_6}{8709120}+\frac{19751 E_2^2 E_4^2}{348364800}-\frac{25117 E_2 E_4 E_6}{1437004800}-}
\eqn{-\frac{225727 E_4^3}{49816166400}-\frac{17009 E_6^2}{8172964800},}
\eq{\wc_{10031}=\wc_{10032}=
-\frac{835 E_2^6}{4478976}-\frac{1181 E_2^4 E_4}{7464960}+\frac{2879 E_2^3 E_6}{19595520}+\frac{17777 E_2^2 E_4^2}{87091200}+\frac{26581 E_2 E_4 E_6}{359251200}-}
\eqn{-\frac{22403 E_4^3}{574801920}-\frac{11161 E_6^2}{269438400},}
\eq{\wc_{22131}=\wc_{22132}=
-\frac{9847 E_2^6}{8957952}+\frac{14387 E_2^4 E_4}{14929920}+\frac{677 E_2^3 E_6}{1632960}-\frac{108697 E_2^2 E_4^2}{522547200}-\frac{3743 E_2 E_4 E_6}{29937600}+}
\eqn{+\frac{99991 E_4^3}{3448811520}+\frac{563 E_6^2}{22453200},}
\eq{\wc_{33331}=\wc_{33332}=
\frac{9889 E_2^6}{8957952}+\frac{13619 E_2^4 E_4}{14929920}-\frac{739 E_2^3 E_6}{19595520}-\frac{127 E_2^2 E_4^2}{34836480}-\frac{2173 E_2 E_4 E_6}{71850240}+}
\eqn{+\frac{1175 E_4^3}{996323328}+\frac{5099 E_6^2}{2451889440},}
\eq{\wc_{20221}=
\frac{1651 E_2^6}{29859840}-\frac{97 E_2^4 E_4}{3317760}-\frac{911 E_2^3 E_6}{19595520}-\frac{157 E_2^2 E_4^2}{41803776}+\frac{37699 E_2 E_4 E_6}{359251200}-}
\eqn{-\frac{51217 E_4^3}{1379524608}-\frac{1679 E_6^2}{38491200},}
\eq{\wc_{33221}=
\frac{38581 E_2^6}{26873856}-\frac{1933 E_2^4 E_4}{8957952}-\frac{9635 E_2^3 E_6}{11757312}-\frac{52331 E_2^2 E_4^2}{104509440}+\frac{16459 E_2 E_4 E_6}{215550720}+}
\eqn{+\frac{11293 E_4^3}{689762304}+\frac{353 E_6^2}{46189440},}
\eq{\wc_{44421}=-\frac{48629 E_2^6}{33592320}-\frac{8813 E_2^4 E_4}{5598720}-\frac{24457 E_2^3 E_6}{117573120}-\frac{28733 E_2^2 E_4^2}{130636800}+\frac{5407 E_2 E_4 E_6}{195955200}+}
\eqn{+\frac{1381 E_4^3}{509483520}+\frac{E_6^2}{1705860}.}


\bibliographystyle{unsrt}
\bibliography{references}

\end{document}